\documentclass[amsmath,10pt,twocolumn,superscriptaddress,footinbib,pre]{revtex4-1}
\usepackage[utf8]{inputenc}
\usepackage{hyperref}
\usepackage{graphicx} 
\usepackage{amsmath,amsfonts,bm} 
\usepackage{uri}
\usepackage{multirow}
\usepackage{booktabs}

\usepackage[x11names, rgb, html]{xcolor}
\definecolor{sbase03}{HTML}{002B36}
\definecolor{sbase02}{HTML}{073642}
\definecolor{sbase01}{HTML}{586E75}
\definecolor{sbase00}{HTML}{657B83}
\definecolor{sbase0}{HTML}{839496}
\definecolor{sbase1}{HTML}{93A1A1}
\definecolor{sbase2}{HTML}{EEE8D5}
\definecolor{sbase3}{HTML}{FDF6E3}
\definecolor{syellow}{HTML}{B58900}
\definecolor{sorange}{HTML}{CB4B16}
\definecolor{sred}{HTML}{DC322F}
\definecolor{smagenta}{HTML}{D33682}
\definecolor{sviolet}{HTML}{6C71C4}
\definecolor{sblue}{HTML}{268BD2}
\definecolor{scyan}{HTML}{2AA198}
\definecolor{sgreen}{HTML}{859900}
\hypersetup{
  colorlinks = true,
  citecolor = {sred},
  urlcolor = {sblue}
}

\begin{document}
\title{Sterically Driven Current Reversal in a Model Molecular Motor}
\author{Alex Albaugh}
\affiliation{Department of Chemical Engineering \& Materials Science, Wayne State University, 5050 Anthony Wayne Drive, Detroit, Michigan 48202, USA}
\affiliation{Department of Chemistry, Northwestern University, 2145 Sheridan Road, Evanston, Illinois 60208, USA}
\author{Geyao Gu}
\author{Todd R.~Gingrich}
\affiliation{Department of Chemistry, Northwestern University, 2145 Sheridan Road, Evanston, Illinois 60208, USA}

\begin{abstract}
Simulations can help unravel the complicated ways in which molecular structure determines function.
Here, we use molecular simulations to show how slight alterations of a molecular motor's structure can cause the motor's typical dynamical behavior to reverse directions.
Inspired by autonomous synthetic catenane motors, we study the molecular dynamics of a minimal motor model, consisting of a shuttling ring that moves along a track containing interspersed binding sites and catalytic sites.
The binding sites attract the shuttling ring while the catalytic sites speed up a reaction between molecular species, which can be thought of as fuel and waste.
When that fuel and waste are held in a nonequilibrium steady-state concentration, the free energy from the reaction drives directed motion of the shuttling ring along the track.
Using this model and nonequilibrium molecular dynamics, we show that the shuttling ring's direction can be reversed by simply adjusting the spacing between binding and catalytic sites on the track. 
We present a steric mechanism behind the current reversal, supported by kinetic measurements from the simulations. 
These results demonstrate how molecular simulation can guide future development of artificial molecular motors.
\end{abstract}

\maketitle

Molecular motors generate directed motion by harnessing free energy gradients, harvested, for example, from the hydrolysis of adenosine triphosphate (ATP) into adenosine diphosphate (ADP) and inorganic phosphate (\( \mathrm{P_{i}} \))
~\cite{brown2017toward,kolomeisky2007molecular,mugnai2020theoretical}.
That directed motion has an essential biological function\textemdash dynein and kinesin transport molecular cargoes on microtubules~\cite{schnapp1989dynein,howard1989movement}, myosin walks along the actin to drive muscle contraction~\cite{finer1994single}, and ATP synthase links rotary motion to chemical synthesis~\cite{noji1997direct}.
The fuel consumption enables directional motion, but the thermodynamic driving force does not fully determine the direction of that motion.
Most myosins hydrolyze ATP to walk along an actin track from - to +, but myosin VI uses the same thermodynamic drive to walk in the opposite direction~\cite{bryant2007power}.
This feature gives myosin VI unique and important biological function~\cite{buss2004myosin}, so significant effort has been devoted to understanding the structural basis for the reversed motion~\cite{liao2009engineered}.
Similar studies have examined the structural basis for directionality in kinesin~\cite{sablin1998direction} and dynein~\cite{can2019directionality}.

Though recent breakthroughs in synthetic chemistry have led to the first artificial autonomous chemically-fueled molecular motors~\cite{wilson2016autonomous,borsley2021doubly,borsley2022autonomous}, it remains a challenge to achieve a similar level of directional control in those designed motors.
The motion of non-autonomous machines driven by magnetic fields~\cite{kline2005catalytic,solovev2010magnetic} can be flipped by inverting the field, but inverting the driving force is less desirable for a chemically fueled autonomous machine.
As in the myosin examples, even without altering the fuel and its thermodynamic driving force, it should be possible to introduce structural changes to move in the opposite direction.
Engineering those structural changes is challenged by the fluctuations exhibited in nanoscale motion~\cite{bustamante2001physics,astumian2007design}, which prevent molecular motors from executing deterministic mechanisms like their macroscopic counterparts.
We therefore must differentiate between three types of reversed motion: (i) a fluctuation that causes a motor of a fixed design to spontaneously move in opposition to its typical behavior~\cite{brown2019theory}, (ii) a reversal of the typical behavior of a fixed design due to a negated thermodynamic drive~\cite{lacoste2008fluctuation}, and (iii) a reversal of the typical behavior by altering a motor design without altering the drive.
Here, we report this final type of reversal in a simulation model~\cite{albaugh2022simulating} inspired by the catenane motor of Wilson et al.~\cite{wilson2016autonomous}.

That catenane motor, the first experimental realization of an autonomous synthetic chemically fueled motor, consists of two interlocked rings.
The larger of the two rings can be viewed as a track around which the smaller randomly diffuses.
Wilson et al.\ showed that by coupling the supramolecular complex to a chemical reaction (the conversion of 9-fluorenylmethoxycarbonyl chloride into dibenzofulvene and carbon dioxide), the diffusion of the smaller shuttling ring can be gated to statistically prefer clockwise motion.
Two of us used that essential design to construct a molecular dynamics model of a catenane-like motor that is similarly coupled to a fuel decomposition~\cite{albaugh2022simulating} .
Like the experimental system, that model consisted of two repeated copies of a motif around the large ring.
These motifs are composed of a binding site that attracts the shuttling ring and a catalytic site where the decomposition reaction is catalyzed.
By introducing grand canonical Monte Carlo (GCMC) moves into the simulation, a nonequilibrium steady state (NESS) is sustained in which fuel is typically added, a decomposition reaction is catalyzed through an interaction with the motor, and the waste products are removed.
The NESS presents the motor with a replenishing supply of high-free-energy fuel whose decomposition can be coupled to directed motion.

While holding fixed the fuel's driving force, we here show that the direction of the motor is reversed by increasing the number of repeated motifs around a shuttling ring of a fixed size.
Upon translating the rotary motor into a linear one, we demonstrate that the reversal arises from the spacing between binding and catalytic sites within the repeated motifs.
The shuttling ring at a binding site sterically repels fuel molecules from accessing nearby catalytic sites, and the kinetics of shuttling ring motion flips depending on whether the steric repulsion occludes the nearest catalytic site or the two nearest sites.
We present that argument at a schematic level, and validate it by measuring the rate of key kinetic steps from the NESS simulations.
In addition to informing the design of experimentally realized catenane motors, our results emphasize how sensitively dynamic function can depend on molecular design.

\section*{Results}
\subsection*{A Rotary Motor}

\begin{figure*}
\centering
\includegraphics[width=0.95\textwidth]{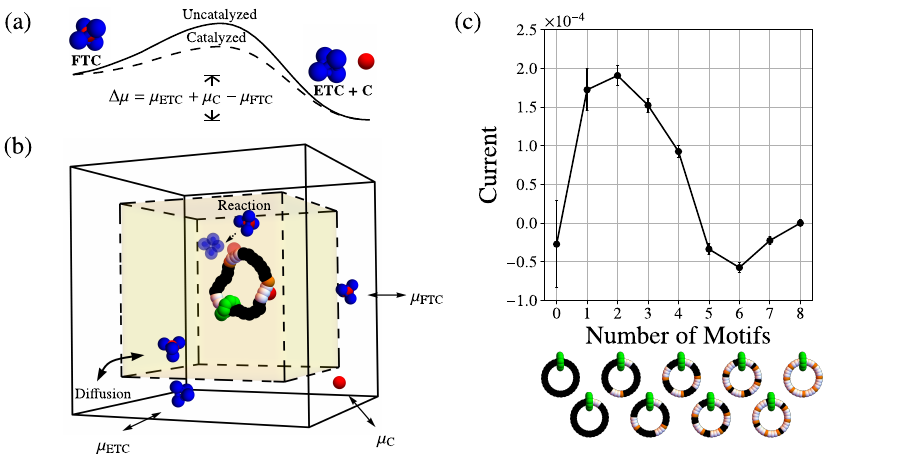}
\caption{
(\textit{a}) The decomposition of a full tetrahedral cluster (FTC) into an empty tetrahedral cluster (ETC) and free central particle (C) provides a thermodynamic driving force to power a molecular motor.
(\textit{b}) That rotary motor comprises a small shuttling ring (green) that diffuses around a larger ring.
The larger ring has repeated motifs consisting of a site that binds the shuttling ring (orange particle) and an adjacent catalytic site (a cluster of three neighboring white particles) that catalyzes the FTC decomposition.
As described in Ref.~\cite{albaugh2022simulating}, the motor is simulated under nonequilibrium steady state (NESS) conditions associated with a surplus of FTC.
A periodic-boundary-condition simulation box is divided into a yellow motor-containing region, modeled by Langevin dynamics, and an exterior white region which supplements the Langevin dynamics with grand canonical Monte Carlo chemostats to hold FTC, ETC, and C species at the respective chemical potentials, \(\mu_{\rm FTC}, \mu_{\rm ETC},\) and \(\mu_{\rm C}\).
Provided diffusion between the two regions is sufficiently fast, the motor's dynamics sample a NESS with a thermodynamic driving force generated by those chemostats.
(\textit{c})  While fixing the large ring to have 32 particles, the number of repeated motifs was varied from zero to eight, with volume-excluding, inert black particles separating those motifs.
The average current (net clockwise hops per time) reverses from positive to negative as more motifs are added.
Data points and error bars are given by the mean and standard error across 100 independent simulations with \(2\times 10^8\) time steps of size \(\Delta t = 5 \times 10^{-3}\).
}
\label{fig:circular}
\end{figure*}

The catenane motors upon which this work is built consist of two motifs on opposite sides of a large ring~\cite{wilson2016autonomous, albaugh2022simulating}.
Those motifs contain adjacent components that bind a shuttling ring and catalyze a reaction.
The regions between motifs are inert, and are represented as volume-excluding particles that serve as an essentially featureless track.
Along that track the shuttling ring diffuses, but the blocking and unblocking of the catalytic sites generates directionality by gating the diffusion in a manner that has been interpreted as an information ratchet~\cite{astumian2016running}.
One may therefore view each motif like a tooth on a ratchet, offering a means to lock in the shuttling ring's forward progress.
That picture suggests that adding more teeth would make it easier to prevent incremental progress from backsliding, thereby amplifying the current, e.g., the rate of net clockwise hops.
We therefore hypothesized that increasing the number of motifs while maintaining the size of the large ring would make a better motor that pushes the shuttling ring in a preferred clockwise direction more efficiently with higher current. 
A previous study showed that creating [2]-catenanes with more than two shuttling ring binding sites is possible~\cite{leigh2003unidirectional}, making this an experimentally realizable configuration.
Fig.~\ref{fig:circular} shows our intuition about teeth on a ratchet was spectacularly wrong.
There, we show results from NESS simulations~\cite{albaugh2022simulating} with a varying number of the motifs, reflecting that the addition of motifs can actually induce current to reverse directions.

Briefly, the simulations introduce a classical cluster of Lennard-Jones particles we call a full tetrahedral cluster (FTC) which, over the course of Langevin dynamics, can decompose into an empty tetrahedral cluster (ETC) and a central particle (C) that had been stuck inside the full cluster~\cite{albaugh2020estimating}.
This reversible decomposition reaction is held away from equilibrium by three separate GCMC chemostats, one each for FTC, ETC, and C.
As pictured, the motifs, spaced as evenly as possible around the large ring, each consist of a single orange particle acting as a binding site next to three white particles acting as a catalytic site.
Details of the attractions and repulsions between particles are discussed in Materials and Methods.
The interlocked rings undergo translational and rotational diffusion within the simulation box, but we focus on the diffusion of the shuttling ring along the track of the large ring.
We measure that current by monitoring which particle along the large ring is closest to the center of mass of the shuttling ring and by recording when the shuttling ring hops clockwise (CW) or counterclockwise (CCW) by one particle.
Integrating those hops over a simulation with \(N_{\rm steps}\) and time step \(\Delta t\) gives a net number of clockwise hops \(\Delta n\) observed in time \(\tau \equiv N_{\rm steps} \Delta t\).
A net hops-per-time current is thus reported as \(j = \Delta n / \tau\).

Here, we considered a ring with \(32\) total particles, so anywhere between zero and eight motifs could fit around the ring.
A motor with either zero or eight motifs is symmetric, requiring the current to vanish in both cases.
In between the extremes, Fig.~\ref{fig:circular}c shows the current rises and falls\textemdash the original 2-motif catenane design moves CW while a motor with 5, 6, or 7 motifs gives CCW currents. 

\subsection*{A Linear Motor}

\begin{figure}
\centering
\includegraphics[width=0.45\textwidth]{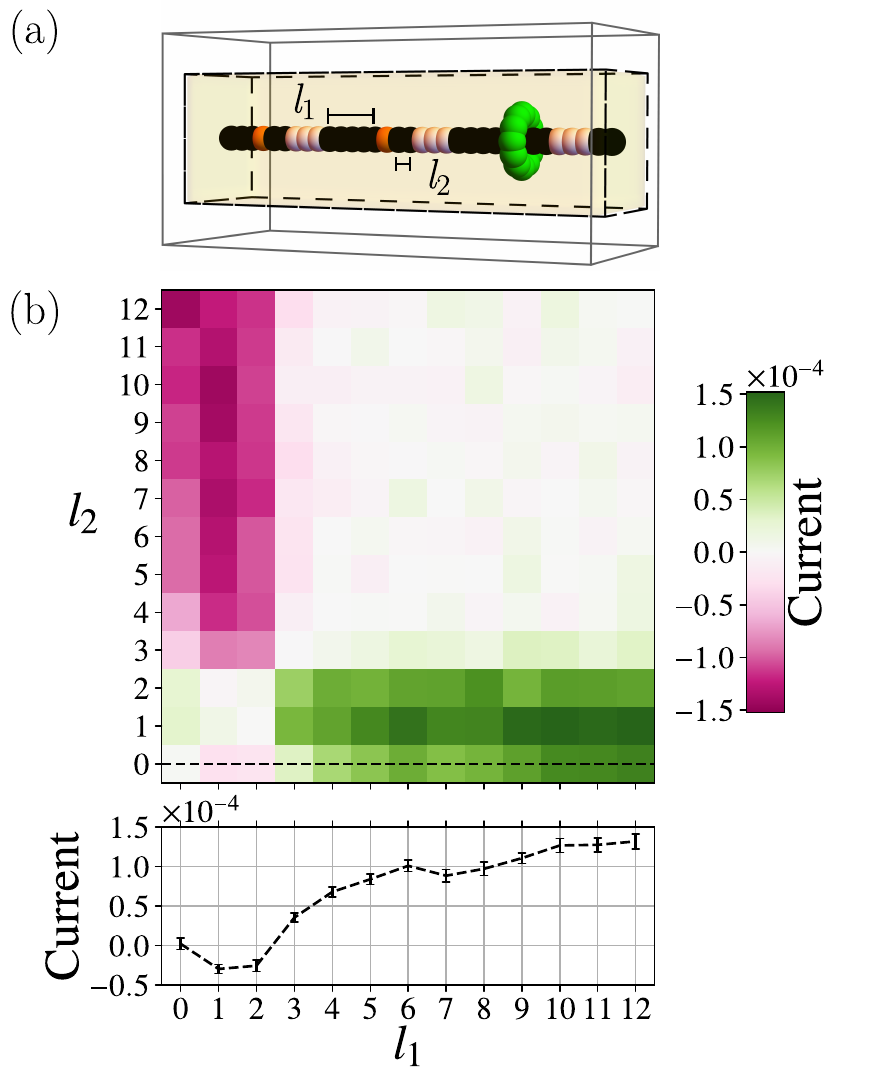}
\caption{
(\textit{a}) The rotary motor of Fig.~\ref{fig:circular} is unfurled into a linear motor with periodic boundary conditions and with orange binding sites separated from white catalytic sites by \(l_1\) and \(l_2\) inert black particles.
The box width was varied along with \(l_1\) and \(l_2\) to simulate a fixed track containing three motifs.
Chemical potentials were imposed in the white region of the simulation box as in Fig.~\ref{fig:circular}.
(\textit{b}) The steady-state current as a function of \(l_1\) and \(l_2\) was measured from 100 independent trajectories with \(2\times 10^8\) time steps of size \(\Delta t = 5 \times 10^{-3}\).
Current is trivially inverted by interchanging \(l_1\) and \(l_2\), but it is also nontrivially inverted along the slice with \(l_2 = 0\) (dashed black line), consistent with the current reversal in Fig.~\ref{fig:circular}. 
Data along that \(l_2 = 0\) slice is highlighted below the heat map to illustrate the standard errors.
}
\label{fig:l1_l2}
\end{figure}

Changing the number of motifs affects not only the number of teeth for a ratchet but also the spacing between a binding site and the catalytic site to the CCW direction.
To study the importance of that spacing, we transformed the rotary motor of Fig.~\ref{fig:circular}b into the linear motor of Fig.~\ref{fig:l1_l2}a, with a dynamic shuttling ring and a fixed periodic track.
Such a linear configuration was suggested in Ref.~\cite{wang2019track}.
Because of the periodic boundary conditions, our linear motor is still effectively a catenane.
The linear geometry, however, eliminates curvature effects while allowing us to systematically change the spacings between adjacent binding and catalytic sites by varying the number of inert (black) particles to the left \((l_1)\) and right \((l_2)\) of each binding site.
Fig.~\ref{fig:l1_l2}b shows a heat map of the current for a range of \(l_1\) and \(l_2\) spacings.
Consistent with the rotary motor of Fig.~\ref{fig:circular}, the direction depends on \(l_1\) when \(l_2 =0\).
Small \(l_1\) gives CCW current while large \(l_1\) flips to a CW current.
By reproducing the current reversal in the linear model, we emphasize that the mechanism for the reversal arises from the spacings between catalytic and binding sites, not the number of motifs.
This fact is further corroborated by SI Fig.~\ref{fig:long}, where we show that the current of an \(l_2 = 0\) linear motor is unaffected by the number of motifs.

\subsection*{A Steric Mechanism for Current Reversal}

\begin{figure*}
\centering
\includegraphics[width=0.95\textwidth]{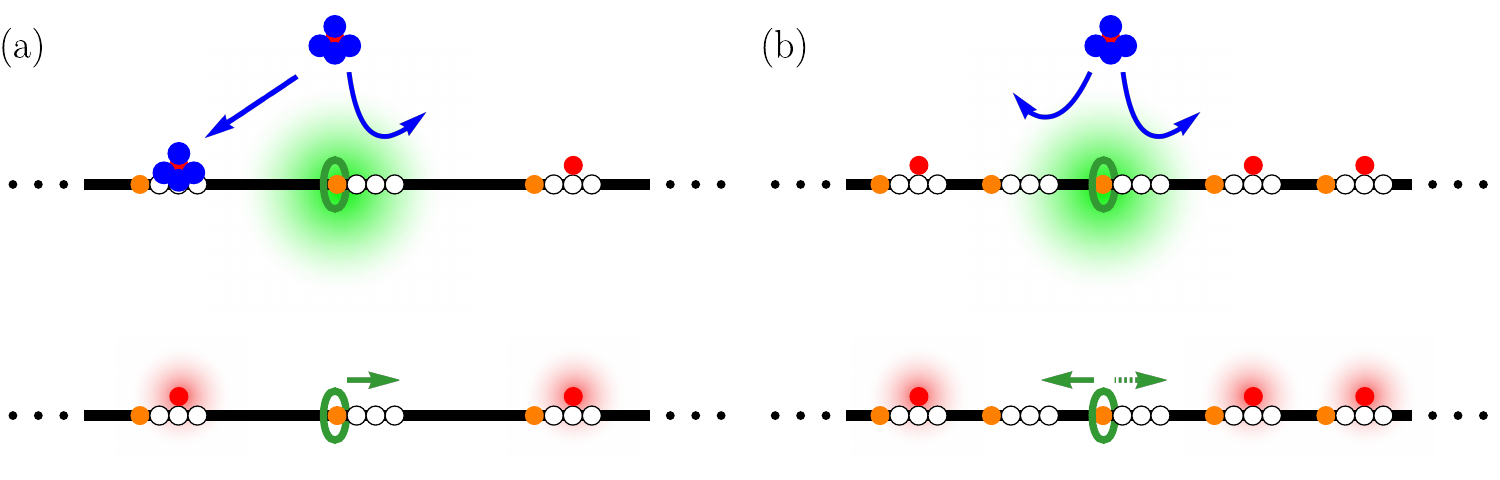}
\caption{
The current reversal originates from a trade-off between two steric effects as the spacing \(l_1\) decreases while \(l_2 = 0\).
(\textit{Top Row})
Generation of blocking groups at catalytic sites is dominated by the attack of an FTC, the decomposition of which can leave behind a bound C.
Steric repulsion from the shuttling ring (shaded green) shields either one (\textit{a}) or two (\textit{b}) catalytic sites from that attack depending on the \(l_{1}\) spacing between motifs.
Consequently, typical NESS configurations (bottom row) mirror the pattern of which sites are accessible to FTC (top row).
In fact, the NESS's strong driving force from FTC to ETC + C imparts an arrow of time, implying that the configurations of the top row flow down into configurations of the bottom row much more than the reversed process.
(\textit{Bottom Row})
(\textit{a}) At large \(l_1\), the shuttling ring tends to be blocked from moving to the left, resulting in rightward motion.
(\textit{b}) At small \(l_1\), the shuttling ring can move either right or left, but the binding site to the right is comparatively unstable due to a second steric effect between the shuttling ring and a bound C (shaded red).
}
\label{fig:linear}
\end{figure*}

That the current reversal occurs when both \(l_{1}\) and \(l_{2}\) are small, hints that it originates from a localized steric effect.
The shuttling ring diffuses along the track but can be impeded when a C blocks the path by binding strongly to a catalytic site.
The directionality of that gated diffusion thus depends on whether a C particle will tend to block the catalytic site to the left or right of the shuttling ring and whether the ring will stably rest at the neighboring binding sites.
As we illustrate in Fig.~\ref{fig:linear}, these effects can act in opposite directions, yielding the current reversal when \(l_{2} = 0\).

To understand that effect, first recognize that C blocks the shuttling ring by binding tightly to the catalytic site, but it can sometimes unbind to let the ring pass.
Because the NESS operates under high-fuel, low-waste conditions, any C that randomly unbinds, is very likely to diffuse away from the motor and be extracted from the system.
To re-block the catalytic site typically requires an FTC to decompose and leave a new C behind.
Thus, the probability that a catalytic site is blocked depends sensitively on whether the site is accessible for an FTC to approach.
The top row of Fig.~\ref{fig:linear} shows that steric repulsions between the shuttling ring and FTC can limit that accessibility on one side of the ring when \(l_1\) is large (Fig.~\ref{fig:linear}a) or on both sides when \(l_1\) is small (Fig.~\ref{fig:linear}b).
Since the generation of blocking groups arises from FTC decomposition, the typical blocking group configurations (bottom row of Fig.~\ref{fig:linear}) mirror the FTC accessibility.
The positive current generated by the large \(l_1\) motor in Fig.~\ref{fig:linear}a straightforwardly follows from those typical blocked configurations; the shuttling ring can move to the right but not the left.
The lower panel of Fig.~\ref{fig:l1_l2}b also shows that the current does not plateau for large \(l_{1}\), but continues to increase.
This increase of the current at large \(l_1\) arises due to the increasing distance between binding sites.
The rate of net jumps between binding sites saturates at large \(l_{1}\), but the current increases since the binding sites are further apart (see SI Fig.~\ref{fig:curr_vs_jumps}).

The situation becomes more nuanced when \(l_1\) approaches the length scale of steric repulsions between FTC and the shuttling ring.
In that case, catalytic sites to both the left and right of the shuttling ring can both be unblocked.
Though the shuttling ring can access the binding sites to the right and left, those sites are not equally stable because of the steric repulsions between the shuttling ring and blocking groups.
The bottom row of Fig.~\ref{fig:linear}b shows that those repulsions (red shaded region) will destabilize the rightward binding site more than the one on the left.
Even if the ring were to make equal attempts to move left and right, the moves to the left would stick more often, yielding the net leftward motion.

We emphasize that the mechanisms for current in both directions depends on the nonequilibrium driving force.
Were it not for the surplus of FTC, the system would obey detailed balance.
FTC would still decompose into ETC and C, with C able to act as a blocking group at a catalytic site, but in equilibrium, it would be equally likely to see ETC and C coalesce into an FTC.
That balance in the chemical reaction translates into a time-reversal symmetry for the shuttling ring\textemdash the forward and reversed trajectories are equally likely.
The thermodynamic force from the driven FTC \(\rightleftharpoons\) ETC + C reaction breaks that symmetry.
Due to the excess of FTC, C typically approaches the ring via one mechanism (extraction of C from FTC at a catalytic site) and departs via a different mechanism (loss of a solitary C with no tetrahedron nearby).
When C is added and removed by a single time-reversed mechanism, the rates of those addition and removal events are coupled together; a faster binding rate also leads to a slower unbinding.
With two mechanisms, however, fuel can be consumed to speed up the binding via the FTC decomposition without slowing the unbinding mechanism.
The result for Fig.~\ref{fig:linear} is that configurations in the top row are rapidly pushed toward the bottom row without a balanced flow from the bottom row back up to the top row, allowing us to reason through the steps in sequence, i.e., first FTC approaches, then C is extracted to bind, then a ring moves.
That the driving force can break time-reversal symmetry to induce a shuttling-ring current is well known~\cite{astumian2016running}; the remarkable feature of the present model is that the sign of that current can flip due to subtle steric effects. 

\subsection*{Other Steric Effects}
While we have highlighted the current-reversal phenomenon as the most striking consequence of steric effects, other notable consequences manifest in Fig.~\ref{fig:l1_l2}.
By symmetry, current must invert upon exchanging \(l_1\) and \(l_2\), so Fig.~\ref{fig:l1_l2} is antisymmetric about \(l_1 = l_2\).
Without loss of generality, we focus on the data with \(l_1 > l_2\).
A prominent feature of Fig.~\ref{fig:l1_l2} is that the current plummets when \(l_2 \geq 3\), a length scale that corresponds to the effective range of repulsions between shuttling ring and ETC or FTC (see SI Fig.~\ref{fig:lj_en_frc}).
If a binding site is more than three particle radii from a catalytic site, then the bound shuttling ring's dynamics becomes decoupled from catalytic decompositions that occur beyond the steric range.

A more subtle feature of Fig.~\ref{fig:l1_l2} is that at fixed large \(l_1\), the current depends nonmonotonically on the \(l_2\) spacing.
We just discussed why current drops when \(l_2\) grows large relative to the shuttling ring's steric repulsions, but the data also show degrading performance when \(l_2\) is made too small.
Specifically, a larger current comes from the intermediate choice \(l_2 = 1\) than from either \(l_2 = 0\) or \(l_2 = 2\).
The schematic in the bottom row of Fig.~\ref{fig:linear}a, which depicts such a large-\(l_1\), \(l_2 = 0\) regime, is useful to understand the cost of making \(l_2\) too small.
For the shuttling ring to move rightward to the empty binding site, it will clash with the bound C, suggesting greater current with larger \(l_2\).
As in the current-reversal phenomenon, the non-monotonicity of current for \(l_2 = 0, 1, 2\) reflects a trade-off between competing effects.
To maximize current, the spacing \(l_2\) should be set to an intermediate value which is large enough but not too large.

The sensitivity of these ``goldilocks'' tradeoffs is especially stark in the regime with both \(l_1 < 3\) and \(l_2 < 3\).
Within that regime, we see current reversal for \(l_2 = 0\).
Why does the current reversal disappear if \(l_2\) is increased by one particle?
That change slightly reduces the impact of the repulsion between shuttling ring and bound C that was highlighted in Fig.~\ref{fig:linear}b.
Because multiple steric effects act in opposition to each other, the net effect can be subtle and challenging to anticipate a priori.
Indeed, our simulations of toy models are most appealing for their ability to highlight the tradeoffs that lead to distinct dynamics behaviors, not necessarily to anticipate a precise value of \(l_1\) and \(l_2\) that would induce an experimental current inversion.
In similar ways, hard particles have been simulated to understand the qualitative structure of liquid crystal phase diagrams though the simplifying model could never be expected to yield precise transition temperatures~\cite{allen1993simulations}.

\subsection*{Validation from NESS Simulations}

\begin{figure*}
\centering
\includegraphics[width=0.75\textwidth]{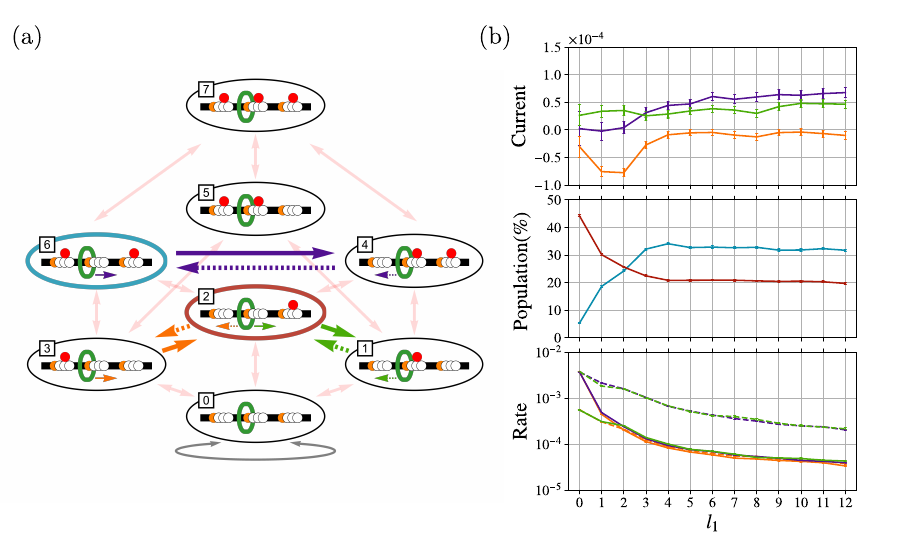}
\caption{
(\textit{a})  Simulated dynamics of the linear motor with three motifs can be coarse-grained onto an 8-state Markov model.
The presence or absence of a blocking group at each of the three catalytic sites define the eight states, and two of the most influential states are highlighted inside blue and red ovals.
Transitions between states occur when a blocking group is added or removed (light red), when the shuttling ring moves on an unblocked track (gray), or when the shuttling ring moves on a partially blocked track (bold purple, orange, and green).
As the generators of current, those purple, orange, and green transitions are highlighted, with solid arrows when the shuttling ring moves to the right (positive) and dashed arrows when it moves to the left (negative).
(\textit{b})  Selected data from the Markov model as a function of the \(l_{1}\) spacing when \(l_{2}=0\).
(\textit{top})  The shuttling-ring current (\(c_{ij}(l_{1}+l_{2}+4)\)) due to currents along the purple (\(c_{46}\)), orange (\(c_{32}\)), and green (\(c_{21}\)) edges.
Summing these three contributions yields the total shuttling-ring current shown in Fig.~\ref{fig:l1_l2}(\textit{b}, \textit{bottom}).
(\textit{middle})  The empirical populations \(p_2\) and \(p_6\) for occupying states 2 (red) and 6 (blue).
(\textit{bottom})  The rates of the highlighted transitions (\(k_{ij}\)) depicted in (\textit{a}) with corresponding colors and line styles.
Data for the states and transitions that are not highlighted is presented in SI Figs.~\ref{fig:full_markov} and~\ref{fig:everything}.
Data points and error bars are given by the mean and standard error across 100 independent simulations with \(2\times 10^8\) time steps of size \(\Delta t = 5 \times 10^{-3}\).
}
\label{fig:markov}
\end{figure*}

To more quantitatively evaluate the current-reversal mechanism, we coarse-grained NESS simulations for a \(l_2 = 0\) linear motor into a kinetic model.
We classified each microstate in the simulation into a mesostate based only on the presence or absence of a blocking group at each of the three catalytic sites, as depicted in Fig.~\ref{fig:markov}.
From these coarse-grained trajectories, we extracted the steady-state probability of each mesostate and the probability per unit time of transitioning between connected states, repeating the calculations for each \(l_1\) spacing.
With those measurements, we fit an \(l_1\)-dependent continuous-time Markov model for the mesostate kinetics.
The Markov model serves as a concise way to consolidate mesoscopic kinetic measurements from NESS simulations, thereby illuminating how mesostate populations, rates, and currents change as \(l_1\) is varied.

All those populations, rates, and currents are reported in SI Fig.~\ref{fig:full_markov}.
We highlight in Fig.~\ref{fig:markov} those data most significant to quantitatively illustrate the mechanism of Fig.~\ref{fig:linear}.
Though the eight mesostates are defined by whether the three catalytic sites are blocked or unblocked, transitions between those states are not only induced by binding and unbinding of a blocking group (light red transitions).
When the shuttling ring moves, it can also induce a change in the mesostate since the identities of the catalytic sites are defined relative to the shuttling ring.
Fig.~\ref{fig:markov} highlights those mesostate transitions that correspond to shuttling ring motion, transitions colored purple, orange, and green.
The net rightward shuttling ring current is the sum of those three currents, oriented in the direction of the solid arrows.
The top of Fig.~\ref{fig:markov}b shows that the current reversal can largely be understood as a transition from dominance of the orange current at small \(l_1\) to the purple current at large \(l_1\).

The positive purple current and negative orange current are both amplified when the population of the originating mesostate increases.
For example, increasing the steady-state probability of the mesostate drawn with a red boundary favors flux along the dashed orange transition and increasing the population of the mesostate drawn with a blue boundary favors flux along the solid purple transition.
The middle of Fig.~\ref{fig:markov}b shows that the steady-state populations shift from the red to blue state as the spacing grows.
That blue mesosate is more prevalent than the red at large \(l_1\) because the high-fuel NESS introduces sufficient FTC to strongly favor configurations with more blocking groups.
For small \(l_1\), however, the population of the red state spikes; steric repulsion between the shuttling ring and the two nearest catalytic sites prevents blocking groups from attaching both to the left and right of the shuttling ring.

These measurements comport with the earlier qualitative steric arguments, placing the orange-to-purple current reversal on more quantitative grounds.
It remains to explain why the negative orange current dominates over the positive green current since a high population of the red state gives a high flux along both the dashed orange and solid green transitions.
As evidenced by the rates in Fig.~\ref{fig:markov}b, there is a notable difference in the orange and green fluxes back into the red state.
The rate of returning to the red state via leftward movement of the shuttling ring (dashed green) is far greater than the rate of returning to the red state via rightward movement of the shuttling ring (orange solid).
That dashed green transition originates from a highly unstable configuration with a shuttling ring adjacent to a blocking group whereas the solid orange transition originates from a state with an extra spacing \(l_1\) between shuttling ring and blocking group.
The overall effect, then, is that leftward movement becomes preferred at small spacing due to destabilizing steric interactions between the shuttling ring and blocking groups.
The currents of Fig.~\ref{fig:markov} were computed by extracting both transition rates and populations from simulation.
It is possible to estimate the current from the rates alone, taking the populations to be the steady state computed from the Markov model.
That analysis, shown in SI Fig.~\ref{fig:compare} generally tracks the measured populations of Fig.~\ref{fig:markov}.

The Markov model analysis serves our purpose of validating the steric argument since the measured rates are useful and sufficient to illustrate kinetic asymmetry~\cite{albaugh2022simulating, penocchio2023kinetic}.
The model is thermodynamically grounded because the microscopic simulation's chemostats explicitly impose the thermodynamic driving force, through chemical potentials for FTC, ETC, and C.
The impact of those chemical potentials on the Markov rates is not transparent, so those rates would need to be re-extracted from simulations when the chemical potentials are altered.
By contrast, when working with the kinetics of elementary reactions, it is possible to simply relate the thermodynamic driving force to rates of transitioning forward and backward along a microscopically reversible elementary reaction~\cite{bauer2014local,astumian2018trajectory,falasco2021local}.
Such a connection between thermodynamics and Markov rates cannot be expected here since the transitions in Fig.~\ref{fig:markov} are coarse grained and combine together all of the different mechanisms to move between states into a single probability per unit time for each step~\cite{esposito2012stochastic}.
We therefore highlight that the Markov model should not be thought of as a model of the elementary kinetic steps.
It should instead be understood as a coarse-grained kinetic model built upon a thermodynamically consistent microscopic simulation.

\section*{Discussion}

We have illustrated a steric mechanism for current reversals of a specific catenane model.
Of course, details of the model will quantitatively impact the transition, altering, for example, the particular distance \(l_1\) at which a current reversal is induced, but our evidence suggests that the basic steric mechanism is very robust.
SI Fig.~\ref{fig:conc} shows the impact of driving the motor more strongly by tuning \(\mu_{\rm FTC}\) so as to induce more FTC fueling reactions to drive directed motion.
For each \(l_1\) spacing, the magnitude of the current grows with fuel concentration until saturation. 
The sign of that current does not depend on \(\mu_{\rm FTC}\), so the current reversal persists whether operating in a weakly or strongly driven regime.
The described simulations use a stochastic integrator to propagate an underdamped Langevin dynamics and SI Fig.~\ref{fig:gamma} furthermore shows that the current reversal is robust across a range of friction coefficients for the Langevin dynamics, strong evidence that the mechanism is not inertial in nature.
A more detailed discussion of damping can be found in SI Sec.~\ref{sec:damping}, where we emphasize that the underdamped integrator can take the motor simulation well into the physically relevant overdamped regime.
Compared to biological motors with Reynolds numbers of roughly \(\mathrm{Re}=10^{-8}\)~\cite{brown2019theory}, we show that the main-text simulations correspond to Re on the order of \(10^{-5}\) and those in SI Fig.~\ref{fig:gamma} approach Re of \(10^{-9}\). 

We are still in the early days of being able to design molecular-scale machines that harvest free energy from their surroundings, particularly autonomous machines that mimic biological motors.
To effectively design such machines we will need to identify design principles and effects that are robust across a range of conditions, such as driving force and friction.
These principles should reveal what is required for molecular motors to approach thermodynamic limits when transducing fuel decomposition into directed motion~\cite{albaugh2023thermodynamic}.
A significant challenge is that the operation of these machines can depend sensitively on the design, a dependence brought into focus by the current reversal.
In light of that sensitivity, it is challenging to systematically engineer molecular motors.
Whereas physicists have studied Brownian ratchet current reversals in response to variations in one or two parameters, e.g., a noise~\cite{doering1994nonequilibrium,kula1998brownian} or driving frequency~\cite{bartussek1994periodically,wickenbrock2011current,strand2020current}, biologists and chemists need ways to handle incredibly high dimensional design spaces.
Protein motors can be mutated in any number of ways.
Unconstrained by natural amino acids, supramolecular machines have even more conceivable variations.

Theory~\cite{amano2021chemical,amano2022insights} and simulation~\cite{korosec2021substrate,courbet2022computational} have important roles to parse the structure-function relationships in those supramolecular machines.
Here, we have shown, through minimal models for NESS simulations, that the direction of a motor can be inverted by adjusting the spacing between binding sites and catalytic sites.
Importantly, that sort of adjustment can be experimentally realized for supramolecular motors like those of Refs.~\cite{wilson2016autonomous} and~\cite{borsley2021doubly} by varying the length of alkane chains separating binding and catalytic sites.
Our simulations are on a coarse-grained scale that does not seek to explicitly represent such atomistic dynamics.
They are, nevertheless, sufficient to capture the steric hindrance, which has been shown to be important in other contexts, such as light-driven machines~\cite{roke2018molecular}.
For these light-driven molecular motors, precise control over the sterics can both increase and decrease the motor's speed~\cite{huber2021steric}.

We anticipate that the described current reversal mechanism will be relevant to many flavors of artificial molecular motors~\cite{erbas2015artificial}, particularly to those applications which require tracks to direct molecular motors~\cite{unksov2022through}.
The rotary motor of Ref.~\cite{wilson2016autonomous} involved a circular track, but in our simulations, the track could be unfurled into a linear track without disturbing the essential behavior.
That linear configuration was introduced here as a means of isolating the origin of the current reversal, but the linear motor can also be useful in its own right.
Whether built from DNA components~\cite{bath2009mechanism,liu2016biomimetic} or supramolecular chemistry, synthetic motors walking along linear tracks could form the basis for nanoscale assembly lines, molecular robots, and active materials~\cite{wang2019track}.
Hopefully, the simulations and mechanisms described here will prove useful in designing those future generations of synthetic motors.

\section*{Methods}

\subsection*{Langevin Dynamics with Chemostats}
As described in more detail SI Section~VI, the motor and its surroundings were simulated with underdamped Langevin dynamics while the concentrations of FTC, ETC, and C were maintained by grand canonical Monte Carlo chemostats~\cite{frenkel2001understanding,gupta2000grand,chempath2003two}.
This merging of Langevin dynamics with chemostats was achieved by confining the motor to an inner box, which was permeable to the FTC, ETC , and C particles but not to the motor~\cite{albaugh2022simulating}.
Fuel and waste diffused in and out of the inner box sufficiently quickly that the concentrations experienced by the motor reflect the chemical potentials imposed by Monte Carlo insertion and deletion moves in an outer box.
This construction for the rotary motor is depicted in Fig.~\ref{fig:circular}b and for the linear motor in Fig.~\ref{fig:l1_l2}a.
Whether in the inner or outer box, the \(i^{\rm th}\) particle with position \(\mathbf{r}_{i}\) and momentum \(\mathbf{p}_{i}\) was evolved in time according to the underdamped Langevin equation:
\begin{equation}
\begin{aligned}
\dot{\mathbf{r}}_{i} & = \frac{\mathbf{p}_{i}}{m_{i}}
\\
\dot{\mathbf{p}}_{i} &= - \frac{\partial U(\mathbf{r})}{\partial \mathbf{r}_{i}} - \frac{\gamma}{m_{i}} \mathbf{p}_{i} + \boldsymbol{\xi}_{i},
\end{aligned}
\label{eq:langevin}
\end{equation}
where the potential energy \(U\) is a function of all positions \(\mathbf{r}\), \(m_{i}\) is the particle's mass, \(\gamma\) is the friction coefficient and \(\boldsymbol{\xi}\) is a white noise satisfying \(\left<\boldsymbol{\xi}_{i}\right> = \mathbf{0}\) and \(\left<\boldsymbol{\xi}_{i}(t) \boldsymbol{\xi}_{j}(t')\right> = 2 \gamma k_{\rm B} T \delta(t - t') \delta_{ij} \mathbf{I}\) at temperature \(T\), where \(\mathbf{I}\) is the identity matrix.
This equation was numerically integrated with a time step \(\Delta t\)~\cite{fass2018quantifying}, described in more detail in SI Section~VI.
All simulations reported in the main text were carried out using non-dimensionalized units with Boltzman constant \(k_{\rm B} = 1\) and with \(T = 0.5, \Delta t = 5 \times 10^{-3}\), and \(\gamma = 0.5\). 
GCMC moves are attempted every 100 Langevin time steps.

\subsection*{Modeling Interactions} 
The motor design is determined by the spacing of various types of beads comprising binding sites, catalytic sites, etc.
The details of the interactions between those beads also strongly impact the behavior of the motor.
Here, we closely followed our prior design of a minimal motor model~\cite{albaugh2022simulating}, with specific modeling details discussed in SI Section~VI. 
Briefly, the motor rings are made up of beads linked together with finitely extensible nonlinear elastic (FENE) bonds, with a three-body angular potential enforcing circular rings.
The tetrahedral clusters, depicted in blue, present in both FTC and ETC are held together with harmonic bonds.
All particles, those in the motor and those in the fuel/waste, additionally interact with all other particles via two-body 12-6 Lennard-Jones interactions, but with coefficients for the 12-term (\(\epsilon_{\rm R}\)) and 6-term (\(\epsilon_{\rm A}\)) independently controlled to separately tune repulsions and attractions, respectively.
Holding fixed the strengths of the bonded interactions, the motor/fuel system is designed by choosing appropriate attraction and repulsion strengths between all pairs of particle types.
Those particle types can be identified from the snapshot images\textemdash different colors are different particle types, but with an important caveat.
The colors actually distinguish the functional role of different particles, not the particle type.
For example, though the four particles that make up a tetrahedral cluster are all colored blue, they represent four distinct particle types (TET1, TET2, TET3, and TET4) with distinct pairwise interaction strengths.
Where possible, we matched all interactions with our prior work~\cite{albaugh2022simulating}, but we made one important modification to the white catalytic site.
In the earlier work, that catalytic site had three distinct particle types, a CAT2 particle bonded to a CAT1 particle bonded to a CAT3 particle, each of which was rendered white.
To ensure that the model would be symmetric when \(l_1\) and \(l_2\) were permuted, it became necessary to re-parameterize a symmetric catalytic site built from a CAT2 particle bonded to a CAT1 bonded to another CAT2, with new pairwise interaction strengths provided in SI Table~\ref{tab:tab1}.

\subsection*{Coarse Graining and Markov State Model}
The continuous-time Markov model of Fig.~\ref{fig:markov}a was parameterized from NESS simulation data.
Coarse graining was performed in two stages: first by mapping each microstate onto one of 24 mesostates (the ``lab frame'') then by clumping together groups of three symmetrically equivalent states that have identical blocking group configurations relative to the position of the shuttling ring (the ``relative frame'').
The coarse graining in the lab frame used mesostates based on the position of the shuttling ring and the presence or absence of a blocking group at each of the three catalytic sites.
To determine whether a catalytic site was blocked or not, we calculated the distance between the middle catalytic particle and all free C particles in the system.
If at least one C particle was within 1.2 distance units of the middle particle of a catalytic site then the site was considered to be blocked.
The shuttling ring location was determined from a core set coarse graining~\cite{buchete2008coarse,schutte2011markov}.
We calculated which linear track particle was closest to the shuttling ring center of mass.
If the closest track particle was a binding site, that binding site was considered to be the current position of the shuttling ring.
Otherwise, the position was considered to be the last visited binding site.
Three catalytic sites can be blocked or unblocked and the ring can reside at one of three possible binding sites, leading to 24 different mesostates in the lab frame.

Taking into account the translational symmetry, it suffices to track the motion in the relative frame measured with respect to the shuttling ring.
Dynamics in the relative frame moves between the 8 states depicted in Fig.~\ref{fig:markov}a.
In building the relative-frame Markov model, we represented the probability of each of the 8 states as a vector \(\mathbf{p}=[p_{0}, p_{1}, p_{2}, p_{3}, p_{4}, p_{5}, p_{6}, p_{7} ]^{T}\), with states numbered as indicated in Fig.~\ref{fig:markov}.
The probability evolves according to the master equation
\begin{equation}
\frac{d \mathbf{p}}{dt} = \mathbf{W} \mathbf{p},
\label{eqn:master_eq}
\end{equation}
where \(\mathbf{W}\) is the continuous-time rate matrix
\begin{equation}
\mathbf{W} = 
\begin{bmatrix}
  -\Sigma_{0}                & k_{10}                                    & k_{20}                                               & k_{30}                                     & 0                                           & 0                                  & 0                                    & 0 \\ 
  k_{01}                         & -\Sigma_{1}                           & k_{21}                                                & 0                                            & k_{41}                                    & k_{51}                          & 0                                   & 0 \\ 
  k_{02}                         & k_{12}                                    & -\Sigma_{2}                                       & k_{32}                                    & k_{42}                                    & 0                                 & k_{62}                             & 0 \\ 
  k_{03}                         & 0                                            & k_{23}                                                & -\Sigma_{3}                           & 0                                           & k_{53}                         & k_{63}                              & 0 \\ 
  0                                 & k_{14}                                    & k_{24}                                                & 0                                            & -\Sigma_{4}                           & 0                                 & k_{64}                             & k_{74} \\ 
  0                                 & k_{15}                                    & 0                                                        & k_{35}                                    & 0                                             & -\Sigma_{5}                & 0                                      & k_{75} \\ 
  0                                 & 0                                            & k_{26}                                                & k_{36}                                    & k_{46}                                     & 0                                 & -\Sigma_{6}                    & k_{76} \\ 
  0                                 & 0                                            & 0                                                        & 0                                            & k_{47}                                     & k_{57}                         & k_{67}                             & -\Sigma_{7}
\end{bmatrix}.
\label{eq:matrixunlumped}
\end{equation}
In choosing this form of the rate matrix, we assume that transitions not drawn in the network of Fig.~\ref{fig:markov} have zero rates.
In reality, the simulated system is soft and very occasionally it may be possible that a shuttling ring squeezes past a blocking group.
Because such transitions were so exceedingly rare, the Markov model treated those events as having zero rates.
We write the total time spent in state \(i\) as \(\tau_i\) and the total number of transitions from \(i\) to \(j\) as \(N_{ij}\), allowing us to estimate the nonzero rates as
\begin{equation}
k_{ij} = \frac{N_{ij}}{\tau_{i}}.
\label{eq:rates}
\end{equation}
Along the diagonal of the rate matrix are the total escape rates from state \(i\), \(\Sigma_i \equiv \sum_{j \neq i} k_{ij}\).
At steady state, \(\frac{d \mathbf{p}}{dt} = \mathbf{0} \) and the steady-state population of states is given by \(\boldsymbol \pi\) equal to the normalized top eigenvector of \(\mathbf{W}\).
The net steady-state probability current along an individual edge connecting states \(i\) and \(j\) is given by 
\begin{equation}
  \label{eq:sscurrents}
\tilde{c}_{ij} = \pi_{i} k_{ij} - \pi_{j} k_{ji}.
\end{equation}
The Markov model's steady-state density can differ slightly from the empirical density (this could happen due to finite-sampling effects, a breakdown of the Markov assumption, or if the disallowed transitions did not truly have zero rates), so we additionally measure the empirical density of each state as the fraction of time spent in that state:
\begin{equation}
p_{i} = \frac{ \tau_{i} } { \tau },
\label{eq:pops}
\end{equation}
where \(\tau = N_{\mathrm{steps}} \Delta t\) is the total simulated time.
Since the simulations provide access to both empirical densities and rate estimates,
We find that current calculations tend to be most robust when we incorporate information from both the empirical densities and the rate estimates (see SI Section~IV).
To do so, we compute the so-called empirical currents
\begin{equation}
  \label{eq:empcurrents}
c_{ij} = p_{i} k_{ij} - p_{j} k_{ji}
\end{equation}
for every edge of the 8-state relative-frame Markov model.

We note that in the relative frame of reference the shuttling ring remains fixed, yet transitions depicted with large purple, orange, green, and gray lines in Fig.~\ref{fig:markov} correspond to transitions where the shuttling ring moves in the lab frame.
We can therefore deduce the lab-frame current from the probability per unit time flowing across those purple, orange, and green edges of the relative-frame Markov model.
Since the shuttling ring will pass \(l_1 + l_2 + 4\) sites upon moving from one binding site to the next, the net shuttling ring current (displacement per unit time) is thus \((l_1 + l_2 + 4) (c_{64} + c_{32} + c_{21})\).
That total current is readily decomposed into the contributions coming from the purple (\(c_{46}\)), orange (\(c_{32}\)), and green (\(c_{21}\)) edges, as illustrated in Fig.~\ref{fig:markov}.
The other transitions in the Markov model (depicted by thin red arrows in Fig.~\ref{fig:markov}) correspond to blocking group creation and removal, with no corresponding shuttling ring movement and therefore do not contribute to the physical movement of the shuttling ring and corresponding current.
Fig.~\ref{fig:markov}b gives data for a select number of states and rates, with the full data set presented in SI Fig.~\ref{fig:full_markov}.

\section*{Acknowledgments}
The authors gratefully acknowledge productive conversations with Rueih-Sheng Fu.
Research reported in this publication was supported by the Gordon and Betty Moore Foundation through Grant No.\ GBMF10790.

\subsection*{Data Availability}
Simulation data, simulation code, and analysis scripts used in this study are available in a public Zenodo.com repository under accession code \url{https://zenodo.org/record/6712829}.

\bibliography{biblio.bib}

\clearpage
\section*{Supporting Information}
\renewcommand{\thefigure}{S\arabic{figure}}
\setcounter{figure}{0}
\renewcommand{\thetable}{S\arabic{table}}
\setcounter{table}{0}

\section{Driving Force}
\label{sec:driving}
\begin{figure}[htb]
\centering
\includegraphics[width=0.33\textwidth]{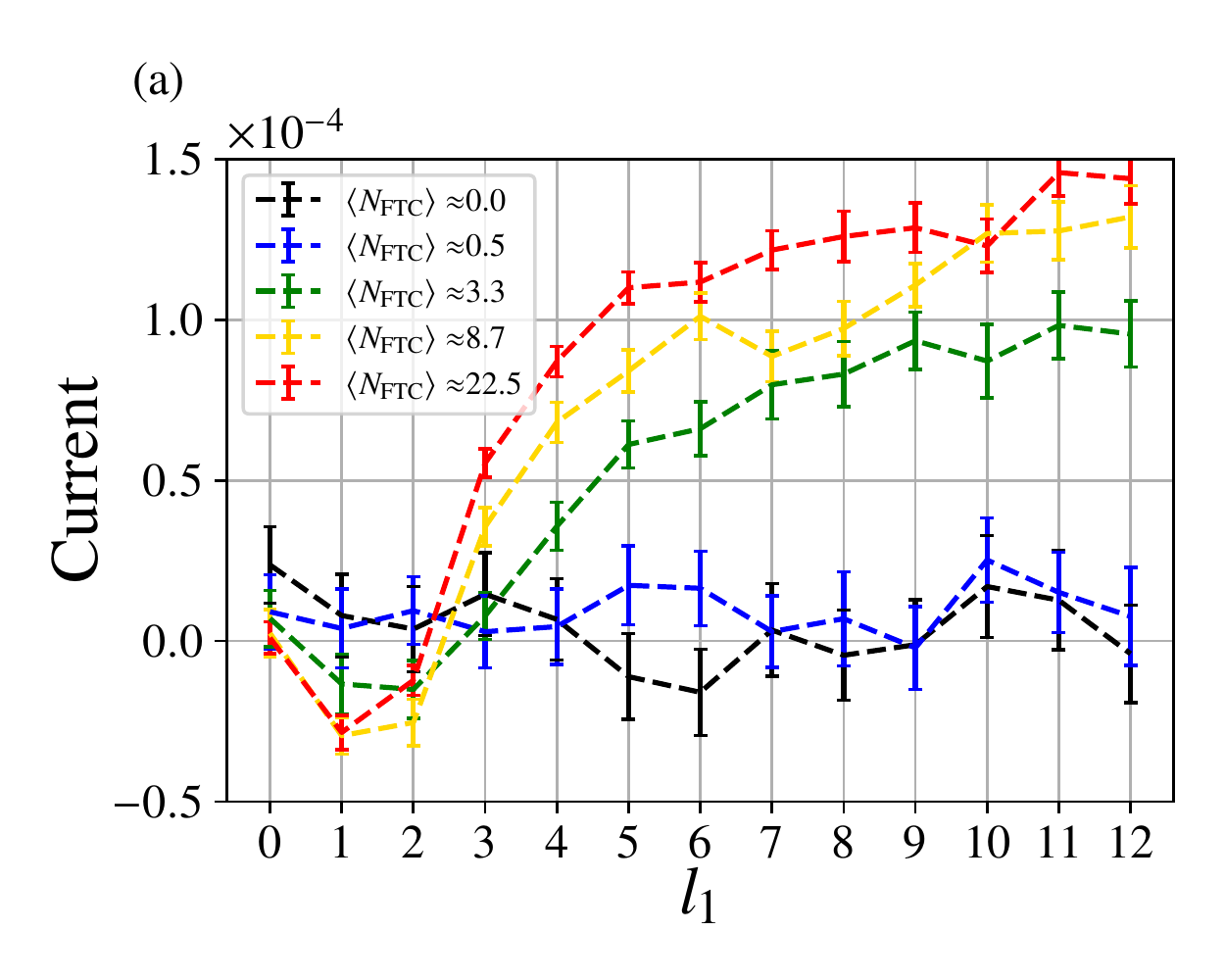}
\includegraphics[width=0.33\textwidth]{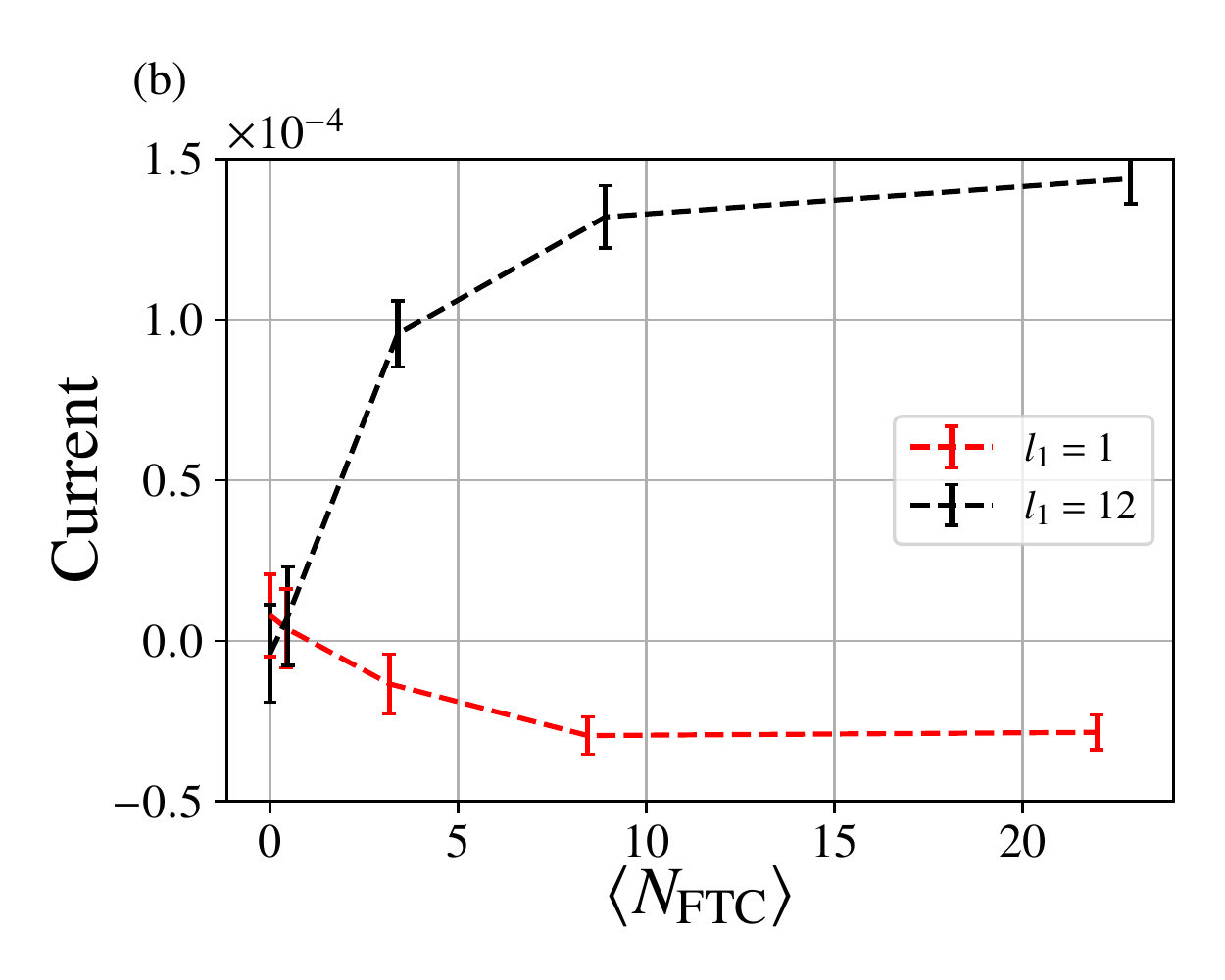}
\caption{
FTC concentration affects motor current magnitude, but not direction.
(\textit{a}) Current as a function of \(l_{1}\) spacing in the linear motor for \(l_{2}=0\) is shown for several values of average number of FTC molecules \(\langle N_{\mathrm{FTC}} \rangle\) in the fixed simulation volume.
(\textit{b}) Current as a function of \(\langle N_{\mathrm{FTC}} \rangle\) for \(l_{1}=1\) (red) and \(l_{1}=12\) (black).
Data points and error bars at \(l_{1}=0\) and \(\langle N_{\mathrm{FTC}} \rangle \approx, 0.0, 0.5, 3.3\) represent  mean and standard error across 200 independent simulations.
All other data points and error bars are given by the mean and standard error across 100 independent simulations.
All simulations have \(2\times 10^8\) time steps of size \(\Delta t = 5 \times 10^{-3}\).
}
\label{fig:conc}
\end{figure}

The concentration of FTC, ETC, and C was controlled by setting the external chemical potentials \(\mu_{\mathrm{FTC}}^{\prime}\), \(\mu_{\mathrm{ETC}}^{\prime}\), and \(\mu_{\mathrm{C}}^{\prime}\), respectively (detailed in SI Sec.~\ref{sec:details}).
As \(l_{1}\) was varied, the simulation cell was contracted or expanded along the perpendicular dimensions so the simulation volume remained fixed.
By maintaining a high concentration of FTC and low concentrations of ETC and C, we placed the motor in a nonequilibrium steady state.
Unless specified otherwise, all simulations in this study used \(\mu_{\mathrm{FTC}}^{\prime}=0.5\) and \(\mu_{\mathrm{ETC}}^{\prime} = \mu_{\mathrm{C}}^{\prime} = -10.0\), which maintained an average of 8.7 FTC molecules, 0.2 ETC molecules, and 1.4 C particles in the simulation volume.
In Fig.~\ref{fig:conc}a we show the motor's current as a function of \(l_{1}\) spacing for several FTC concentrations, spanning both near-to and far-from-equilibrium regimes.
When there is no FTC present the motor is in equilibrium and produces no appreciable current.
Increasing the number of FTC generates more current---negative current at low \(l_{1}\) and positive current at high \(l_{1}\).
While the FTC concentration controls the magnitude of the current, it has no effect on the motor's direction.
After catalytic sites become saturated additional fuel molecules have no effect on the motor's performance, resulting in a plateau in plots of current versus the average number of FTC molecules, \(\langle N_{\mathrm{FTC}} \rangle\).

\section{Periodic Boundary Correlations}
\label{sec:correlations}
\begin{figure}[htb]
\centering
\includegraphics[width=0.33\textwidth]{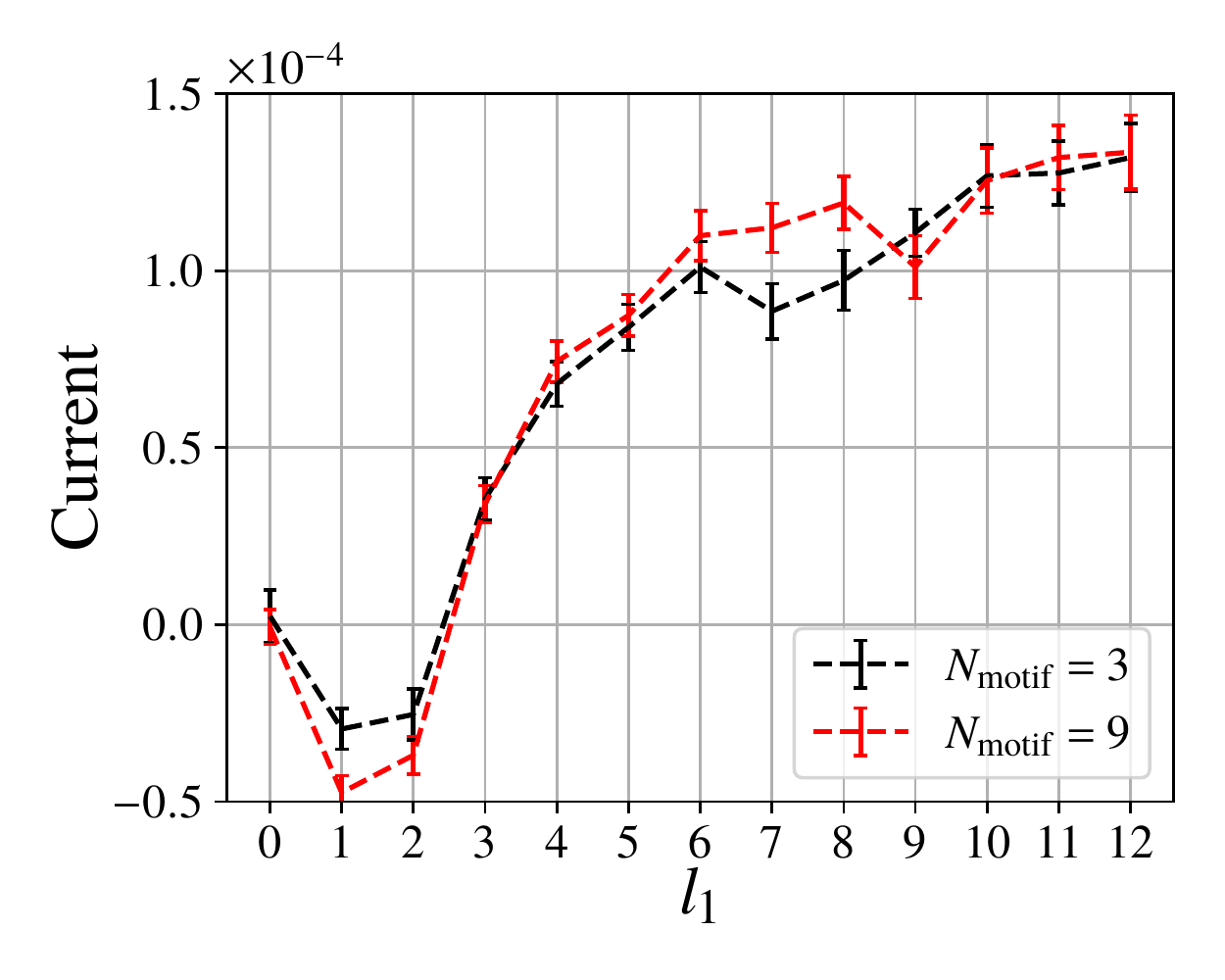}
\caption{
Current reversal persists as the number of motifs \(N_{\mathrm{motif}}\) increases for the linear motor.
For all simulations shown \(l_{2}=0\) as \(l_{1}\) is varied.
Data points and error bars are given by the mean and standard error across 100 independent simulations with \(2\times 10^8\) time steps of size \(\Delta t = 5 \times 10^{-3}\).
}
\label{fig:long}
\end{figure}

We presented two closely related geometries for motors, the catenane rings and the linear motor with periodic boundary conditions.
That linear motor had only three motifs regardless of the spacings \(l_{1}\) and \(l_{2}\).
Due to the periodic boundary conditions, the linear motor's shuttling ring effectively saw more motifs, but the states of those periodic replica motifs were correlated.
One might worry that those correlations would make it problematic to compare, for example, a six-motif ring motor with a three-motif linear motor.
Even if the two geometries had identical spacings, the period boundary condition correlations would differ.  
We confirmed that these correlations did not appreciably affect the results or analyses, as shown in Fig.~\ref{fig:long}.
By increasing the number of explicit motifs in the linear motor from three to nine, and therefore decreasing correlations, the current is not significantly affected and the current reversal is unchanged.

\section{Damping \& Time Scales}
\label{sec:damping}
\begin{figure}
\centering
\includegraphics[width=0.33\textwidth]{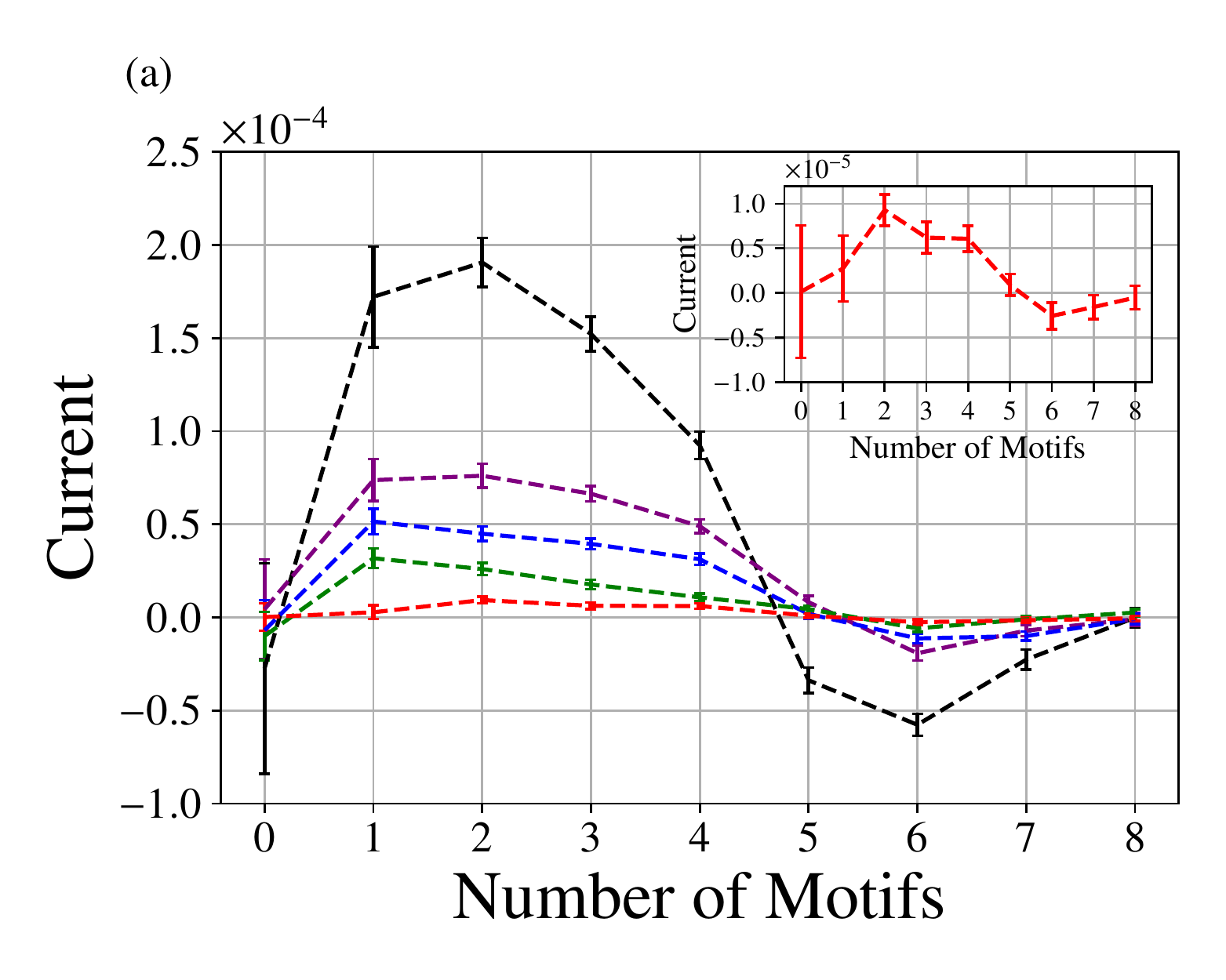}
\includegraphics[width=0.33\textwidth]{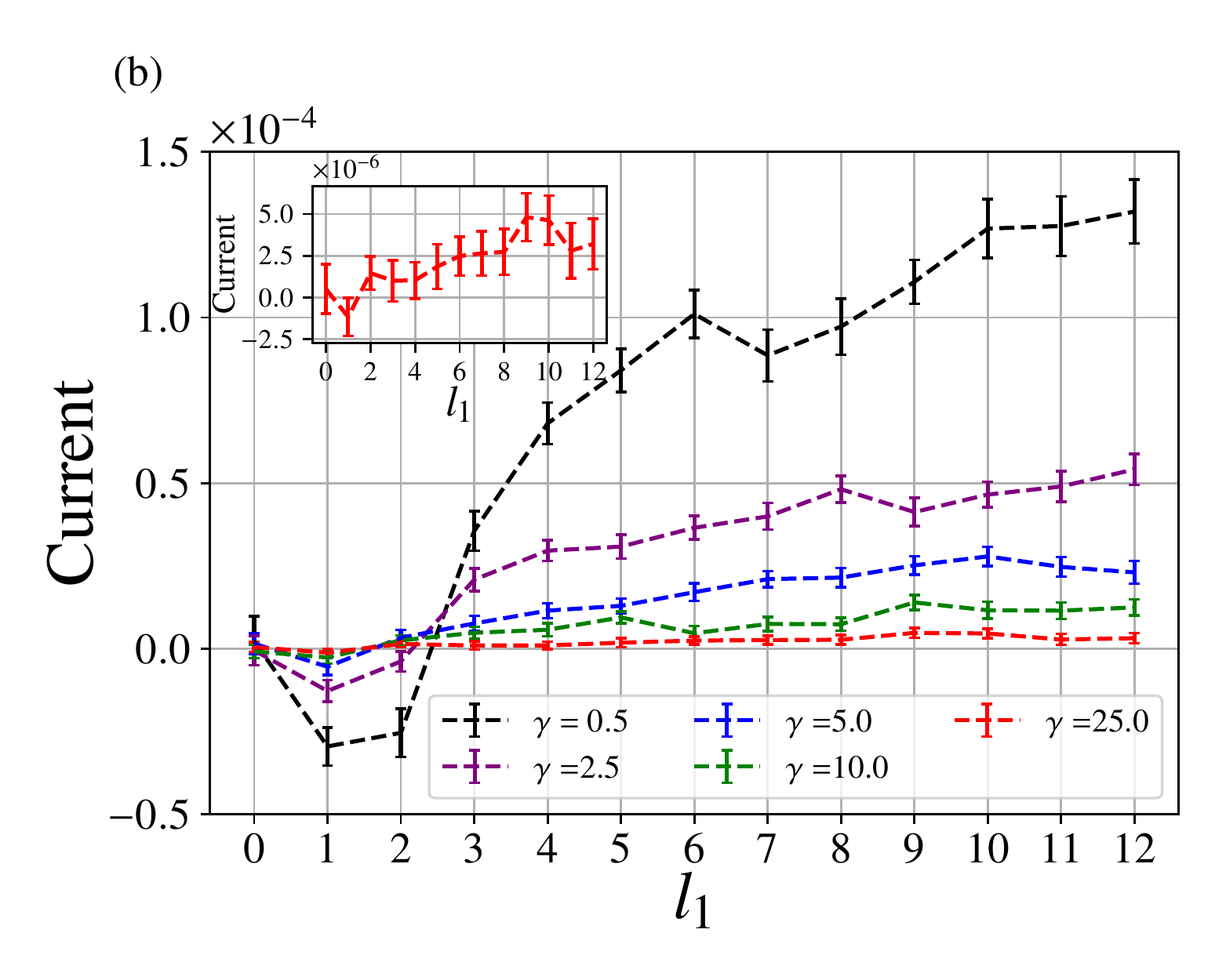}
\caption{
Current reversal persists as the friction of the system \(\gamma\) increases for both the (\textit{a}) rotary motor and (\textit{b}) linear motor.
For the circular motor, the current (given as the net number of particle hops the shuttling ring makes along the track per time) switches from clockwise current (positive) to counterclockwise current (negative) as the number of motifs increases for all values of \(\gamma\).
As in Fig.~\ref{fig:circular}, the circular track is fixed at 32 particles as we vary the number of motifs.
For the linear motor, the \(l_{2}\) spacing is fixed at 0 and the \(l_{1}\) spacing is varied.
The current switches from positive at large spacing to negative a small spacing for all values of \(\gamma\).
Insets show \(\gamma=25\) current data on a smaller scale for ease of visualization.
Data points and error bars for the linear motor at \(\gamma=25.0\) ((b), red curve) represent  mean and standard error across 200 independent simulations.
All other data points and error bars are given by the mean and standard error across 100 independent simulations.
All simulations have \(2\times 10^8\) time steps of size \(\Delta t = 5 \times 10^{-3}\).
}
\label{fig:gamma}
\end{figure}

We also investigated the motor's performance at higher friction coefficients \(\gamma\) to ensure that the cause of the current reversal was not related to inertia.
Many biological motor systems operate in low Reynolds number environments where viscous forces dominate inertia~\cite{astumian2007design}.
Increasing \(\gamma\) creates more particle drag, decreasing the effect of inertia.
The results of these high \(\gamma\) simulations are shown in Fig.~\ref{fig:gamma}a for a circular motor as in Fig.~\ref{fig:circular}b and in Fig.~\ref{fig:gamma}b for a linear motor as in Fig.~\ref{fig:l1_l2}c.
As the friction increases the motor slows down, with the magnitude of current decreasing for all numbers of motifs.
The current reversal observed at a large number of motifs, however, remains present, demonstrating the current reversal is not an inertial effect.

While the motor's shuttling ring moves along a one-dimensional track, we emphasize that the friction coefficient is not the effective damping on a single degree of freedom representing the shuttling ring's location.
Rather, the friction explicitly acts on each particle.
The ring's motion emerges from many interactions of explicitly simulated particles, so it is possible for the ring's motion to appear overdamped if it were to be reduced to a one-dimensional model even if the simulations are run with more modest \(\gamma = 0.5\) damping acting on the particles.
In that manner, some components like a free C particle may have some modest inertial memory even when other components have dynamics that are effectively overdamped.

To parse the impact of the friction coefficient, we determined the time scales of dynamical events (a time step, a bond vibration, free C diffusion, the damping time scale, ETC, FTC, and ring diffusion, transitions in the Markov models, catalyzed and uncatalyzed reactions, and net current) at three different friction coefficients: \(\gamma=0.5\), \(10\), and \(25\).
These time scales, collected in Table~\ref{tab:timescale}, show that most events occur on time scales far longer than the damping time scale.
At \(\gamma=0.5\), free C diffusion and bond vibrations still have mild inertial character, but the functional components of the motor move much more slowly than inertia is lost.

\begin{table*}
\centering
\caption{Time scales for varying \(\gamma\).}
\label{tab:timescale}
\begin{tabular}{  c  r  | c | c | c  }
\multicolumn{2}{ c |}{ } & \(\gamma = 0.5\) & \(\gamma = 10\) & \(\gamma = 25\) \\
\hline
\multirow{12}{1.5em}{\bf{Time Scales}} & time step & \(5 \times 10^{-3}\) & \(5 \times 10^{-3}\) & \(5 \times 10^{-3}\) \\
& bond vibration & \(6 \times 10^{-2}\) & \(6 \times 10^{-2}\) &  \(6 \times 10^{-2}\) \\
& C diffusion &  \(6 \times 10^{-2}\) &  \(10^{0}\) &  \(3 \times 10^{0}\) \\
& damping &  \(2 \times 10^{0}\) &  \(10^{-1}\) & \(4 \times 10^{-2}\)  \\
& ETC diffusion & \(2 \times 10^{0}\)  & \(4 \times 10^{1}\)  & \(10^{2}\)  \\
& FTC diffusion & \(3 \times 10^{0}\)  & \(5 \times 10^{1}\)  & \(10^{2}\)  \\
& ring diffusion & \(10^{3}\)  & \(2 \times 10^{3}\)  & \(3 \times 10^{3}\)  \\
& Markov transitions & \(3 \times 10^{2}\)  - \(3 \times 10^{4}\) & \(10^{2}\) - \(6 \times 10^{5}\) & \(7 \times 10^{1}\) - \(10^{7}\) \\
& uncatalyzed reaction & \(10^{2}\) & \(10^{2}\) & \(2 \times 10^{2}\) \\
& catalyzed reaction & \(3 \times 10^{2}\) & \(2 \times 10^{3}\) & \(5 \times 10^{3}\) \\
& current & \(8 \times 10^{3}\) & \(8 \times 10^{4}\) & \(3 \times 10^{5}\) \\
& simulation & \(10^{6}\) & \(10^{6}\) & \(10^{6}\) \\

\hline

\multirow{2}{1.5em}{\bf{Reynolds Numbers}} & \(c=1\) & \(1 \times 10^{-5}\) &\( 6 \times 10^{-7} \) & \(1 \times 10^{-7}\) \\
& \(c=0.05\) & \(7 \times 10^{-7}\) & \( 3 \times 10^{-8}\) & \(7 \times 10^{-9} \) \\

\hline
\end{tabular}
\end{table*}

Estimates of the hierarchy of time scales in Table~\ref{tab:timescale} came from a variety of analytical and computational methods.
The damping time scale is given as \(m/\gamma\) where \(m\) is an individual particle mass of 1.0 for all particle types~\cite{albaugh2022simulating}.
The bond vibration time scale was estimated as the inverse angular frequency of the harmonic spring connecting two tetrahedral particles \(\sqrt{\mu_{m}/k}\) where \(k=120\) is the force constant and \(\mu_{m}=0.5\) is the reduced mass of the two particle system~\cite{albaugh2022simulating}.
The C, ETC, FTC, and shuttling ring diffusive time scales were estimated by calculating their center of mass velocity autocorrelations and the Green-Kubo relation \(D = \int_{0}^{\infty} dt^{\prime} \langle \mathbf{v}_{\mathrm{com}}(0) \cdot \mathbf{v}_{\mathrm{com}}(t^{\prime}) \rangle\).
The velocity autocorrelations were collecting from simulations of a lone example of each species at \(k_{\mathrm{B}}T = 0.5\) with no GCMC moves and \(2 \times 10^{6}\) time steps with the first \(2 \times 10^{5}\) time steps discarded.
To get the diffusion coefficient we numerically integrated the resulting autocorrelations with a trapezoidal rule with respect to \(t^{\prime}\) from 0 to 50, a time sufficient for all correlations to decay to negligible values.
For FTC the procedure was repeated until a nonreactive trajectory was found.
The shuttling ring was interlocked with a fixed linear track composed of only inert particles (no binding sites) and the diffusion calculation was restricted to the axial direction along the track (\(x\)-direction).
The diffusive time scale was estimated as \(L^{2}/D\) where \(L\) was a characteristic length for each species.
For C we estimated \(L\) as the particle radius 0.45~\cite{albaugh2022simulating}.
For ETC and FTC we estimated \(L\) as the centroid-to-vertex distance of a tetrahedron with a side length equal to two particle radii of 1.0~\cite{albaugh2022simulating}, \(L \approx \sqrt{6}/2\).
For the shuttling ring we estimated \(L\) as the ring radius, where the circumference is approximated by the number of ring particles (12) multiplied by the approximate bond length (1.0), \(L \approx 6/\pi\).
The Markov transition time scales were simply the inverses of the Markov rates described in SI Sec.~\ref{sec:details} and shown in Fig.~\ref{fig:full_markov} for \(\gamma = 0.5\).
For Table~\ref{tab:timescale} we report the values at \(l_{2}=0\) and the minimum and maximum time scales across \(l_{1}=0,1,\dots,12\).
The uncatalyzed and catalyzed reaction time scales are given as the average time between respective reactions where uncatalyzed reactions occur more than 2.0 units away from a catalytic site and catalyzed reactions occur within 2.0 units.
These values were again averaged over all \(l_{1}=0,1,\dots,12\) for \(l_{2}=0\).
The current time scale is the time scale for the shuttling ring to have a net displacement of the size of a single track particle, essentially the inverse current.
We used the current at \(l_{1}=12\) for this time scale for each \(\gamma\).

At \(\gamma=0.5\), damping occurs slower than the diffusive time scale for C, but faster than all other diffusive time scales and, most importantly, faster than any state transitions or currents associated with the motor.
As \(\gamma\) is increased the effect is a slowing of physical time scales like diffusion and motor operation and a quickening of the damping time scale.
At \(\gamma =10.0\) and above even the diffusive motion of C is on a slower scale than the damping and the damping begins to act on the time scale of the fastest motions, the bond vibrations.
Crucially, at all simulated \(\gamma\) values the ring diffusion, reaction, motor transitions, and current occur on time scales much slower than the damping.
While the underlying integration uses an underdamped Langevin method, this does not imply that the resulting dynamics are underdamped.
Using an underdamped Langevin integrator with a large \(\gamma\) can simply be more accurate and efficient than an overdamped Brownian integrator for equivalent dynamics~\cite{ladd2009numerical}, provided the underdamped integrator is \(\gamma\)-limit convergent~\cite{leimkuhler2023contraction}.
Around \(\gamma=0.5\) some bulk diffusive processes may not be fully damped, but across all tested \(\gamma\) values the motor itself is well damped.
The slightly underdamped nature of the bulk at \(\gamma=0.5\) aides in keeping the nonequilibrium concentrations homogenous even though the moderating GCMC moves occur far from the motor toward the boundary of the simulations.
For \(\gamma > 25\) the physical processes occur on time scales too slow to tractably sample through simulation.

To confirm that the motor's motion should be thought of as overdamped, we estimated the Reynolds number of the motor in our simulations.
The Reynolds number is a dimensionless number that quantifies the relative strength of inertial and viscous effects, \(\mathrm{Re} = \rho v L / \mu\).
Here \(\rho\) is the fluid density, \(v\) is velocity, \(L\) is a characteristic length scale, and \(\mu\) is the fluid viscosity. 
For our motors the velocity is the current (the net number of particles hops of the shuttling ring per time) and we selected the current at \(l_{2}=0\) and \(l_{1}=12\) as a characteristic example.
As before, we estimated the characteristic length scale as the approximate radius of the shuttling ring \(6/\pi\).
We approximated the viscosity using the Stokes-Einstein relation \(\mu = k_{\mathrm{B}}T/(6 \pi D L)\).
As we are using a Langevin equation to approximate a solvent, there is no explicitly defined fluid density.
We can set this density as some factor \(c\) of the particle density of the shuttling ring particles \(3m / (4 \pi \sigma^{3})\) where \(m=1\) is the shuttling ring particle mass and \(\sigma=1.0\) is the shuttling ring Lennard-Jones radius~\cite{albaugh2022simulating}.
Conservatively, we can set \(c=1\), stating that the fluid and particle densities are equal.
Studies of the Langevin equation, however, have suggested that it is valid in regimes where the fluid density is less than that of the explicit solute density, i.e. \(c\approx0.05\)~\cite{heyes1999calculation}.
Reynolds numbers in each case and for several \(\gamma\) values are reported in Table~\ref{tab:timescale}.
In all cases, even at \(\gamma=0.5\), we see that \(\mathrm{Re} \ll 1\), which implies that viscous forces dominate inertial forces and the motor is overdamped.
Biological molecular motors in bulk water are estimated to have \(\mathrm{Re} \approx 10^{-8}\)~\cite{brown2019theory}.
Using values of 1000 kg/\(\mathrm{m}^3\) and \(10^{-3}\) Pa\(\cdot\)s as the density and viscosity of water, a shuttling (benzylic amide) ring radius of 10~\AA, and a current of about 100~\AA~per 12 hours, we estimate that the artificial molecular motor on which we based this model~\cite{wilson2016autonomous} has \( \mathrm{Re} \approx 10^{-16}\).
While our model does not quite reach a biological Reynolds number and is far from that of the artificial molecular motor, it is nonetheless much less than unity, indicating that the motor is well damped and viscosity dominates any inertia. 

In addition to an analysis of the damping, Table~\ref{tab:timescale} confirms important time scale ordering critical to the motor's operation.
First, catalyzed reactions must occur faster than ring movement between binding sites.
Without this, the ring's position would not have a strong effect on the relative catalytic reaction rate between sites close to and far from the ring.
This kinetic asymmetry is how the motor couples reactions to directed motion.
At \(\gamma=0.5\) the catalyzed reaction time scale is faster than even the ring's free diffusion along the track.
This is not true at \(\gamma =10\) and \(\gamma =25\), however the catalyzed reaction time scale is still faster than the time scale for the shuttling ring to move from one binding site to another.
This binding site hopping time scale is given roughly as \(k_{00}^{-1}\) (see SI Sec.~\ref{sec:complete} for details) and is \(9 \times 10^{3}\) - \(4 \times 10^{5}\) and \(2 \times 10^{4}\) - \(10^{7}\) for \(\gamma=10\) and \(\gamma=25\), respectively, compared to the catalyzed reaction time scales of \(2 \times 10^{3}\) and \(5 \times 10^{3}\), respectively.
The motor clearly catalyzes reactions to create blocking groups faster than the shuttling ring moves between sites, allowing the shuttling ring to effectively inhibit blocking group creation nearby and generate current.
Another important ordering is that of the reaction and diffusion.
In order to effectively fuel the motor, the fuel diffusive time scale must be faster than the fuel reactive time scale or else fuel would decay before it ever reached the motor.
This ordering is valid for all \(\gamma\) values studied, but at \(\gamma=25\) the time scales get close with FTC diffusion happening at a time scale of \(10^{2}\) and uncatalyzed reaction happening at a time scale of \(2 \times 10^{2}\).
The difference of time scales at \(\gamma=25\) is much smaller than lower \(\gamma\) values and may be a reason why the current is much reduced at \(\gamma=25\), as a comparatively larger amount of fuel will decay in the bulk before ever reaching the motor.
These time scale orderings match what is expected from the experimental system this model is based on~\cite{wilson2016autonomous}.
Indeed, the model was built to satisfy these time scales so as to effectively generate current.

\begin{table*}
\centering
\caption{Dimensionalized model parameters and data.}
\label{tab:dimension}
\begin{tabular}{  c  r | c | c | c | c  }
& & \multicolumn{2}{ c |}{ \(\gamma = 0.5\)} & \multicolumn{2}{ c }{ \(\gamma = 25\)  }  \\
\hline
\multirow{3}{1.5em}{\bf{Chosen Scales}} & particle radius (nm) & \(1\) & \(10\) & \(1\) & \(10\) \\
& particle density (g/cm\textsuperscript{3}) & \(2\) & \(20\) & \(2\) & \(20\) \\
& temperature (K) & \(275\) & \(300\) & \(275\) & \(300\) \\
\hline
\multirow{3}{1.5em}{\bf{Dimensional Results}} & total time (ms) & \(0.03\) & \(30\) & \(0.03\) & \(30\) \\
& viscosity (mPa\(\cdot\)s)& \(0.01\) & \(0.02\) & \(0.7\) & \(0.8\) \\
& FTC concentration (mM) & \(0.4\) & \(4 \times 10^{-4}\) & \(0.3\) & \(4 \times 10^{-3}\)  \\
& FTC \(\to\) ETC + C rate constant (s\textsuperscript{-1}) & \(2 \times 10^{6}\)  & \(2 \times 10^{3}\) & \(2 \times 10^{6} \) & \(2 \times 10^{3}\) \\
& motor cycling rate (Hz) & \(2 \times 10^{5}\) & 200 & \(2 \times 10^{3}\) & \(6 \) \\
& C binding strength (kJ/mol) & \(30\) & \(30\) & \(30\) & \(30\) \\

\hline
\end{tabular}
\end{table*}

As the model was cast in non-dimensional form, we can set scales to examine how it maps onto physical systems.
Suppose one wanted to construct one of these motors at the scale of colloids, where each simulated ball corresponds to a colloid whose interactions with the other colloids could be tuned, for example, through DNA coatings~\cite{angioletti2016theory}.
By choosing a particle radius, particle density, and temperature we can set the characteristic length scale, mass scale, and energy (time) scale, respectively.
Table~\ref{tab:dimension} shows some of the simulation results dimensionalized for reasonable scaling values for \(\gamma=0.5\) and \(\gamma=25\).
Particle radii of 1 nm and 10 nm represent small and medium sized nanoparticles, respectively.
Particle densities of 2 g/cm\textsuperscript{3} and 20 g/cm\textsuperscript{3} could represent moderate density (silica) and high density (metallic) nanoparticles, respectively.
Temperatures of 275 K and 310 K represent cold water and body temperature, respectively.
The \(\gamma=25\) simulations correspond to a physical situation.  
While the small, light particle/low temperature case  implies somewhat large reaction rates and motor speeds, the large, heavy particle/high temperature case is reasonable in all values.
The viscosities are now comparable to water (about 1.5  mPa\(\cdot\)s at 275 K and  0.7 mPa\(\cdot\)s at 310 K) or other common solvents and the reaction rate constant and motor cycling rate decrease to reasonable values, as well.
In general, we see that the reaction rate constant is quite large when dimensionalized.
This is not necessarily a problem, as it indicates that our fuel is relatively unstable relative to the other time scales in the system.
A more stable fuel would increase the motor's efficiency and as long as the relative time scale ordering is not disturbed (i.e. the catalyzed fuel reactions should still happen faster than ring cycling), the motor will still operate properly.
The dimensionalized parameters for \(\gamma=0.5\) are less relevant if the free C is to be viewed as a colloidal particle.
Using the Stokes-Einstein relation to extract the corresponding viscosity of the \(\gamma=0.5\) implicit solvent gives \(0.01\) mPa\(\cdot\)s, a value that is too low for a reasonable liquid.
Rather than representing a mesoscale complex built from colloids, the \(\gamma=0.5\) simulations are more reflective of situations where free C particles retain some inertia because their length scale approaches that of the solvent molecules.
That the current reversal mechanism persists across the range of \(\gamma\) suggests that it has relevance for both the atomistic supramolecular scale and also mesoscale machines that could be build from colloidal particles.

\section{Extended Discussion of the Markov State Models}
\label{sec:complete}
\begin{figure*}
\centering
\includegraphics[width=0.75\textwidth]{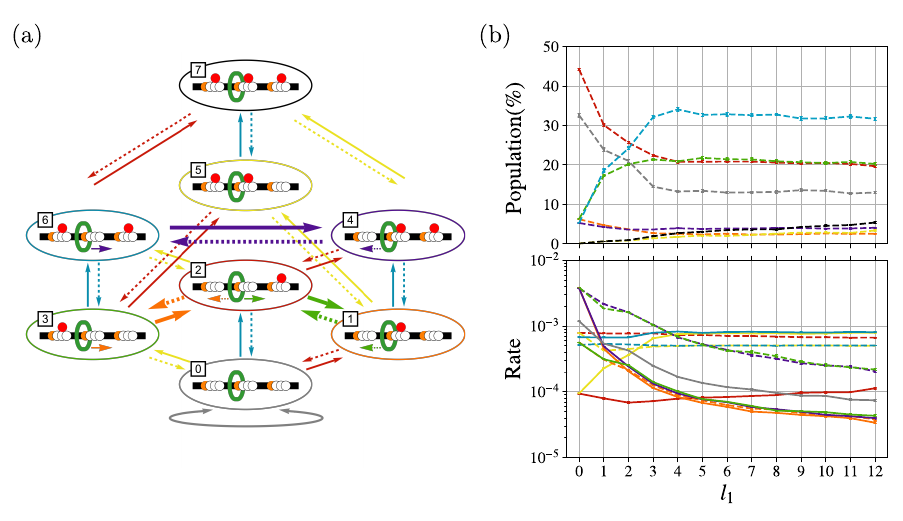}
\caption{
(\textit{a}) The 8-state Markov model of Fig.~\ref{fig:markov} with all states and rates shown.
(\textit{b})  The populations of the Markov states (\textit{top}) and the rates of allowed transitions between states (\textit{bottom}) as a function of the \(l_{1}\) spacing with \(l_{2}=0\). 
The outline color of the state in (\textit{a}) corresponds to its curve color in (\textit{b}) (top) and the arrow color and line style (solid or dashed) for a transition in (\textit{a}) corresponds to the same color and style in (\textit{b}) (bottom).
Data points and error bars are given by the mean and standard error across 100 independent simulations with \(2\times 10^8\) time steps of size \(\Delta t = 5 \times 10^{-3}\).
}
\label{fig:full_markov}
\end{figure*}

\begin{figure}
\centering
\includegraphics[width=0.33\textwidth]{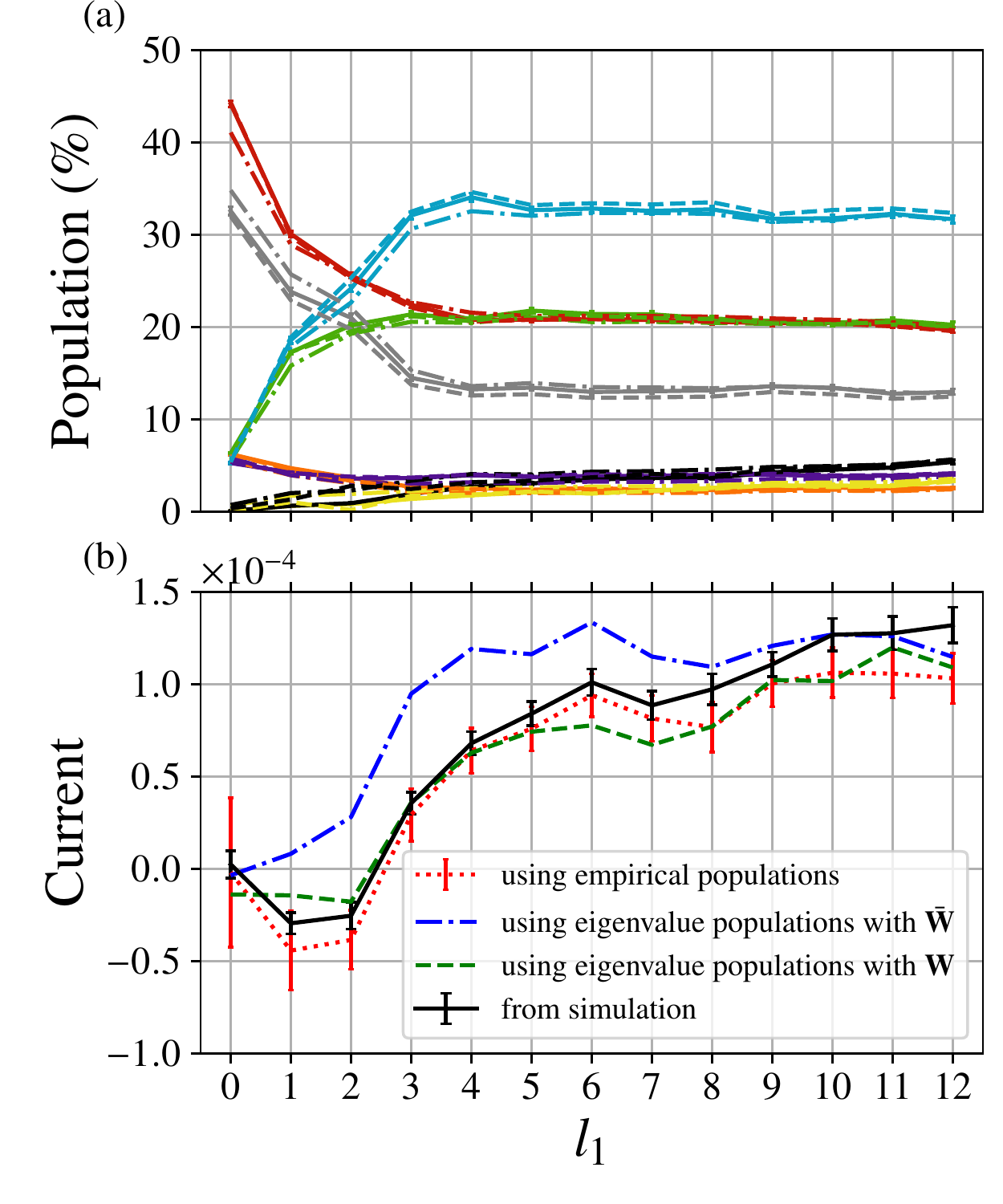}
\caption{
  There is general agreement between the simulation and Markov models.
Simulated empirical populations (\textit{a}) and currents (\textit{b}) (calculated as the net number of hops the shuttling ring makes per time) are plotted with solid lines.
Corresponding data from the lumped-rate Markov model with \(\bar{\mathbf{W}}\) are shown with dot-dashed lines, where the currents are computed using the steady-state probabilities as in Eq.~\eqref{eq:pops}.
By lumping rates together into the Markov model with \(\bar{\mathbf{W}}\), the model deviates at small \(l_1\) (dot-dashed lines).
Those deviations are rectified either by using the more general \(\mathbf{W}\) Markov model from from Eq.~\eqref{eq:rates} of the main text (dashed lines) or by using empirical populations to compute currents across edges of a graph according to Eq.~\eqref{eq:matrix} (dotted lines).
Data points and error bars are given by the mean and standard error across 100 independent simulations with \(2\times 10^8\) time steps of size \(\Delta t = 5 \times 10^{-3}\).
}
\label{fig:compare}
\end{figure}

The data presented in Fig.~\ref{fig:markov}b highlighted only those rates that directly contribute to shuttling ring motion.
The Markov model described in Materials and Methods is parameterized by 24 additional rates involving the addition and removal of blocking groups.
For the purposes of viewing those addition and removal rates as a function of \(l_1\), it is convenient to not have to consider 24 different rates, motivating us to lump together multiple similar addition and removal processes.
By ``lump together'', we mean that we approximate the distinct transition as having identical rates (e.g., \(k_{01} = k_{24} = k_{35} = k_{67}\)).
In Fig.~\ref{fig:full_markov}a, this lumping manifests in multiple red edges describing the binding and unbinding of a blocking group adjacent to the shuttling ring, irrespective of whether other blocking groups are present at more distant sites.
At large \(l_1\) or \(l_2\), the equivalence of these similar transitions is justified, but at small spacings it can start to break down.
If the aim is to generate a Markov State Model that quantitatively reproduces the current for all spacings, the rate matrix in the main text is most appropriate (see Fig.~\ref{fig:compare}).
If, however, the aim is to see how these different attachment rates tend to depend on spacings, we find it very instructive to perform the lumping by forcing the rate matrix to take the form
\begin{equation}
\bar{\mathbf{W}} = 
\begin{bmatrix}
  -\Sigma_{0} & k_{10}                                    & k_{20}                                               & k_{30}                                     & 0                                           & 0                                  & 0                                         & 0 \\ 
  k_{01}                         & -\Sigma_{1} & k_{21}                                               & 0                                             & k_{20}                                   & k_{30}                          & 0                                          & 0 \\ 
  k_{02}                         & k_{12}                                    & -\Sigma_{2} & k_{32}                                    & k_{10}                                    & 0                                 & k_{30}                                   & 0 \\ 
  k_{03}                         & 0                                            & k_{23}                                                & -\Sigma_{3} & 0                                            & k_{10}                         & k_{20}                                   & 0 \\ 
  0                                 & k_{02}                                    & k_{01}                                                & 0                                            & -\Sigma_{4} & 0                                 & k_{64}                                   & k_{30} \\ 
  0                                 & k_{03}                                    & 0                                                        & k_{01}                                    & 0                                             & -\Sigma_{5} & 0                                            & k_{20} \\ 
  0                                 & 0                                            & k_{03}                                                & k_{02}                                    & k_{46}                                     & 0                                 & -\Sigma_{6} & k_{10} \\ 
  0                                 & 0                                            & 0                                                        & 0                                            & k_{03}                                     & k_{02}                         & k_{10}                                     & -\Sigma_{7}
\end{bmatrix},
\label{eq:matrix}
\end{equation}
obtained by setting
\begin{align*}
  k_{03}&=k_{15}=k_{26}=k_{47}=\frac{N_{03} + N_{15} + N_{26} + N_{47} }{\tau_{0} +\tau_{1} +\tau_{2}+\tau_{4}}\\
  k_{30}&=k_{51}=k_{62}=k_{74}=\frac{N_{30} + N_{51} + N_{62} + N_{74} }{\tau_{3} +\tau_{5} +\tau_{6}+\tau_{7}}\\
  k_{02}&=k_{14}=k_{36}=k_{57}=\frac{N_{02} + N_{14} + N_{36} + N_{57} }{\tau_{0} +\tau_{1} +\tau_{3}+\tau_{5}}\\
  k_{20}&=k_{41}=k_{63}=k_{75}=\frac{N_{20} + N_{41} + N_{63} + N_{75} }{\tau_{2} +\tau_{4} +\tau_{6}+\tau_{7}}\\
  k_{01} &=k_{24}=k_{35}=k_{67}=\frac{N_{01} + N_{24} + N_{35} + N_{67} }{\tau_{0} +\tau_{2} +\tau_{3}+\tau_{6}}\\
  k_{10}&=k_{42}=k_{53}=k_{76}=\frac{N_{10} + N_{42} + N_{53} + N_{76} }{\tau_{1} +\tau_{4} +\tau_{5}+\tau_{7}}.
\end{align*}
As mentioned in Materials and Methods, \(N_{ij}\) is the number of transitions from state \(i\) to state \(j\) and \(\tau_i\) is the amount of time spent in state \(i\), both calculated directly from simulated trajectories.
At steady state \(\frac{d \mathbf{p}}{dt} = \mathbf{0} \) and the steady-state population of states is given by the normalized top eigenvector of \(\mathbf{W}\).
We compare the empirical populations from simulation with the eigenvector populations in Fig.~\ref{fig:compare}a.
Those populations are identical at large \(l_{1}\).
For small \(l_{1}\), the equivalence of lumped rates (e.g. \(k_{03}=k_{15}=k_{26}=k_{47}\)) breaks down, accounting for deviations between empirical and top-eigenvector populations.

\begin{figure*}
\centering
\includegraphics[width=0.75\textwidth]{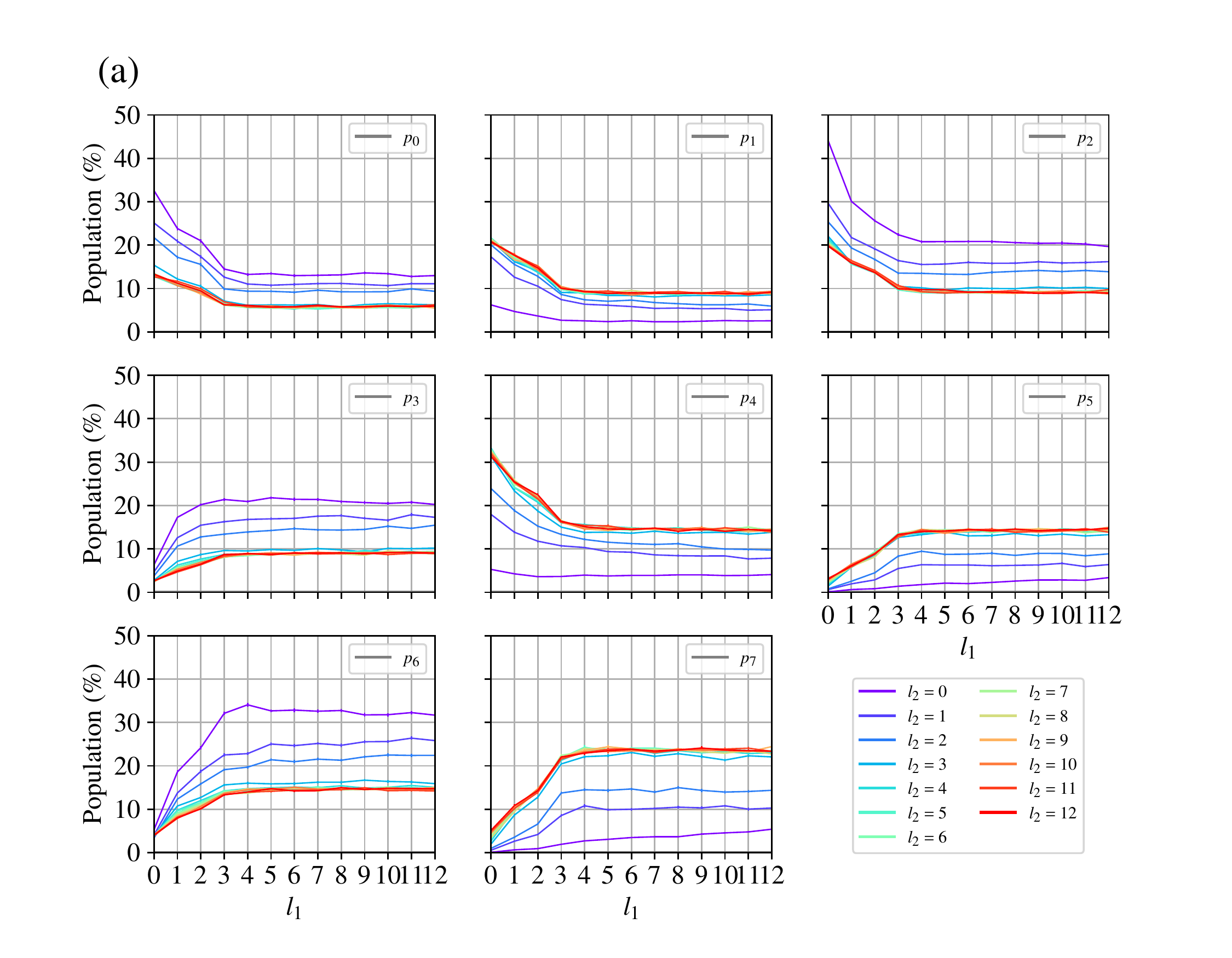}
\includegraphics[width=0.75\textwidth]{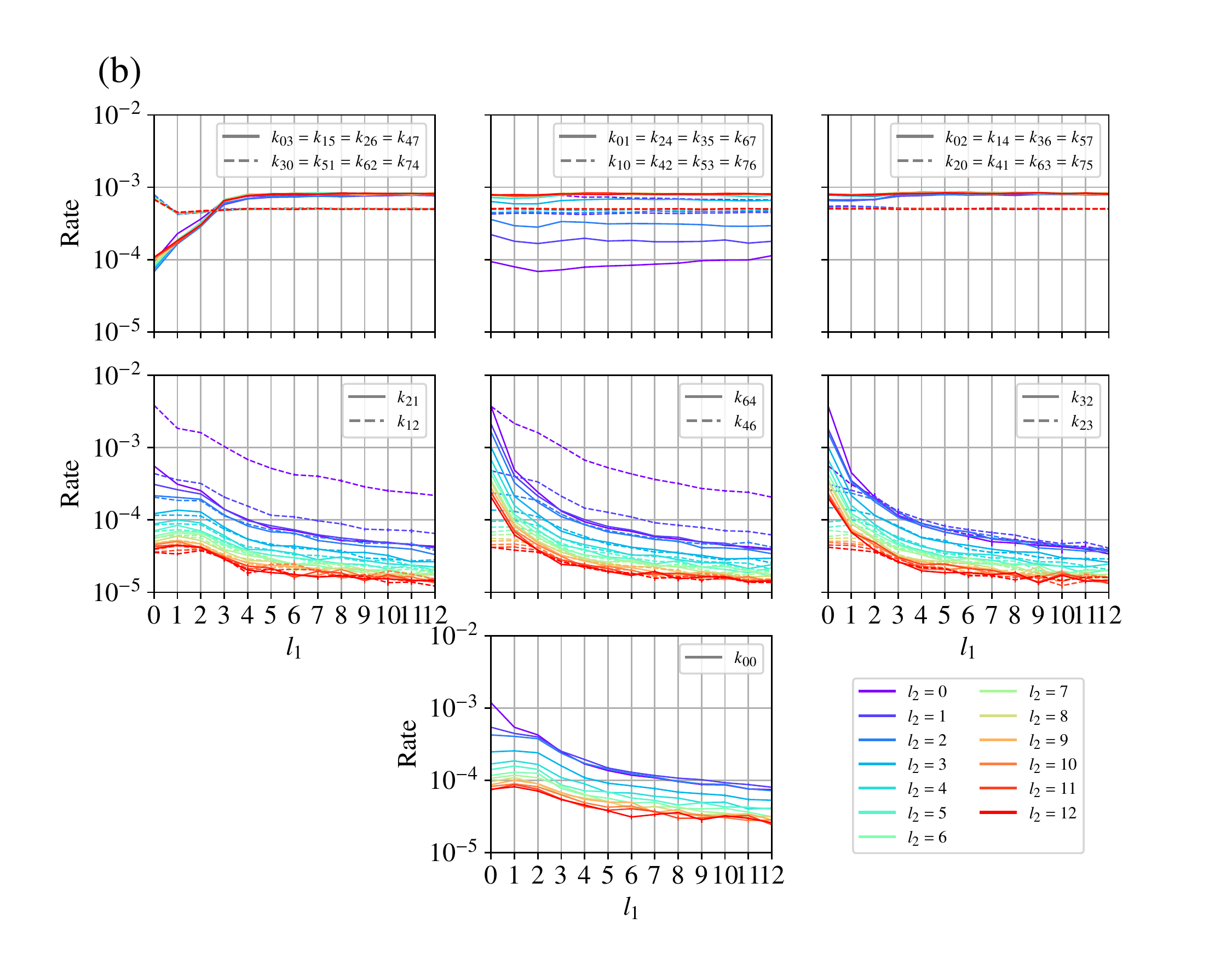}
\caption{
The Markov state populations (\textit{a}) and rates (\textit{b}) for all combinations of \(l_{1}\) and \(l_{2}\) values studied.
Populations and rates are plotted as a function of \(l_{1}\) with different color curves denoting different values of \(l_{2}\).
Where possible rates in one direction along one edge of the Markov graph (Fig.~\ref{fig:full_markov}) are plotted with the rates in the opposite direction along that edge using different line styles (solid and dashed).The population of state \(i\) is labeled as \(p_{i}\) and the transition rate from state \(i\) to state \(j\) is labeled as \(k_{ij}\).
The states \(i\) and \(j\) are numbered from 0 to 7 corresponding to the labels in Fig.~\ref{fig:full_markov}a.
Data points and error bars are given by the mean and standard error across 100 independent simulations with \(2\times 10^8\) time steps of size \(\Delta t = 5 \times 10^{-3}\).
}
\label{fig:everything}
\end{figure*}

Fig.~\ref{fig:full_markov}a shows all the populations and lumped rates for the 8-state relative-frame Markov model with \(l_2 = 0\), the motor that exhibits current reversal, and Fig.~\ref{fig:everything} shows the Markov state populations and rates for all \(l_{1}\) and \(l_{2}\) values.
From Fig.~\ref{fig:everything}, it is clear that populations and rates plateau as \(l_{1}\) and \(l_{2}\) grow large, demonstrating that the interesting changes that occur at small \(l_{1}\) and \(l_{2}\) are due to short-range steric effects.
At large \(l_{1}\) and \(l_{2}\) spacing, the reaction events and shuttling ring position decouple, resulting in no current and rates independent of spacing.

\section{Additional Steric Mechanism Data}
\label{sec:additional}

\begin{figure}
\centering
\includegraphics[width=0.23\textwidth]{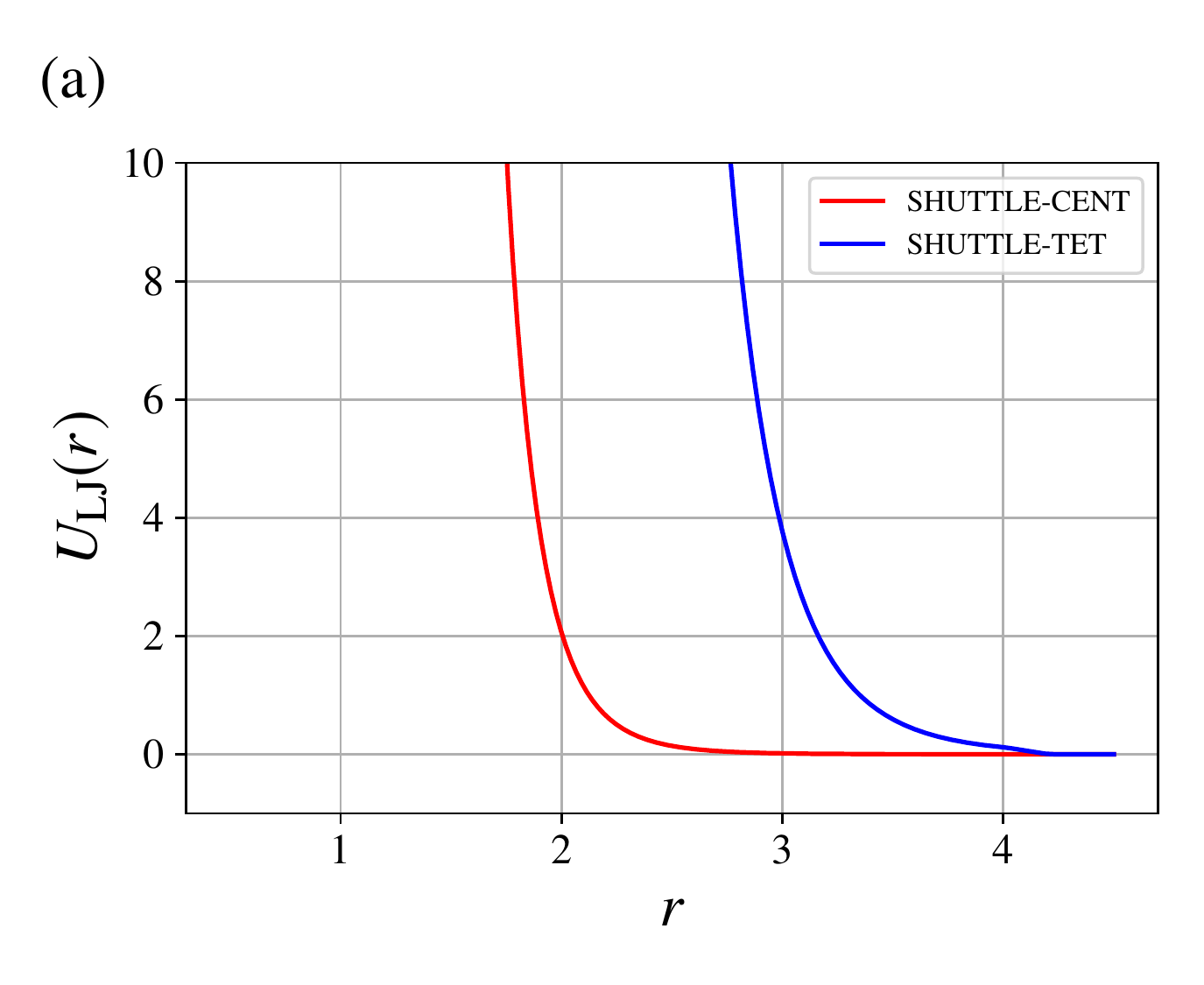}
\includegraphics[width=0.23\textwidth]{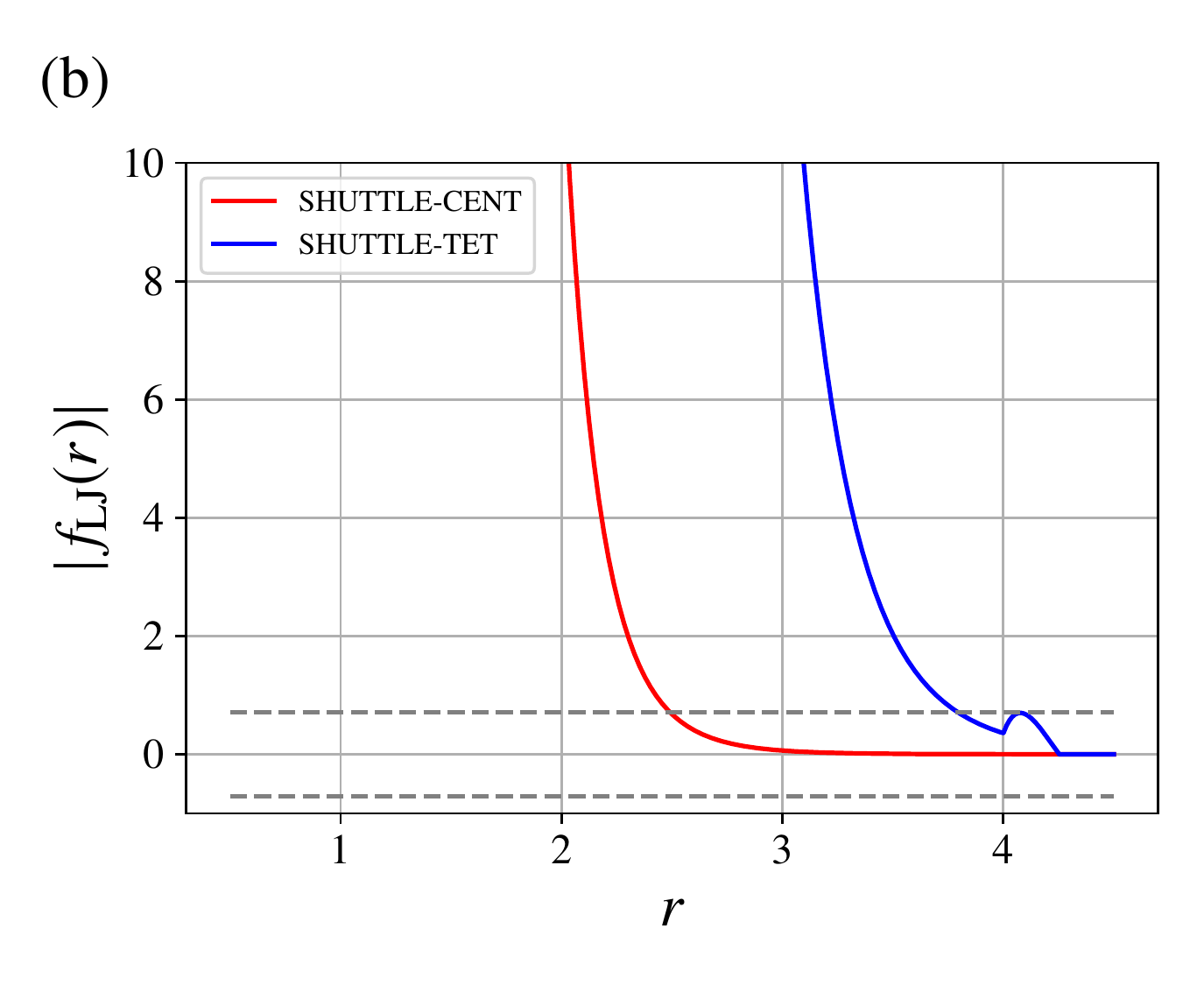}
\caption{
The modified, switched Lennard-Jones energy (\textit{a}) and force (\textit{b}) (given by \eqref{eq:switch}) between a blue tetrahedron particle (TET) and a green shuttling ring particle (SHUTTLE), shown in blue, and between a red C particle (CENT) and a green shuttling ring particle (SHUTTLE), shown in red.
The dashed gray lines in (\textit{b}) represent the standard deviation of the zero-mean random force in the Langevin equation (Eq.~\eqref{eq:langevin}) for comparison.
}
\label{fig:lj_en_frc}
\end{figure}

In understanding the steric mechanisms that contribute to the motor's current reversal, it is important to understand the range of these steric interactions.
Fig.~\ref{fig:lj_en_frc} shows both the interaction energy and force between a blue particle (part of ETC and FTC clusters) and a green particle (part of the shuttling ring) and also the interaction between a red particle (C) and a green particle.
See Figs.~\ref{fig:circular}a,~\ref{fig:circular}b, and~\ref{fig:l1_l2}b for depictions of these particle types.
From Fig.~\ref{fig:lj_en_frc} we can see that the effective steric range between the shuttling ring and FTC/ETC is greater than the effective range between the shuttling ring and C.
These distances are approximately 3 and 2 distance units, respectively, or about 3 and 2 particle radii.

\begin{figure}
\centering
\includegraphics[width=0.33\textwidth]{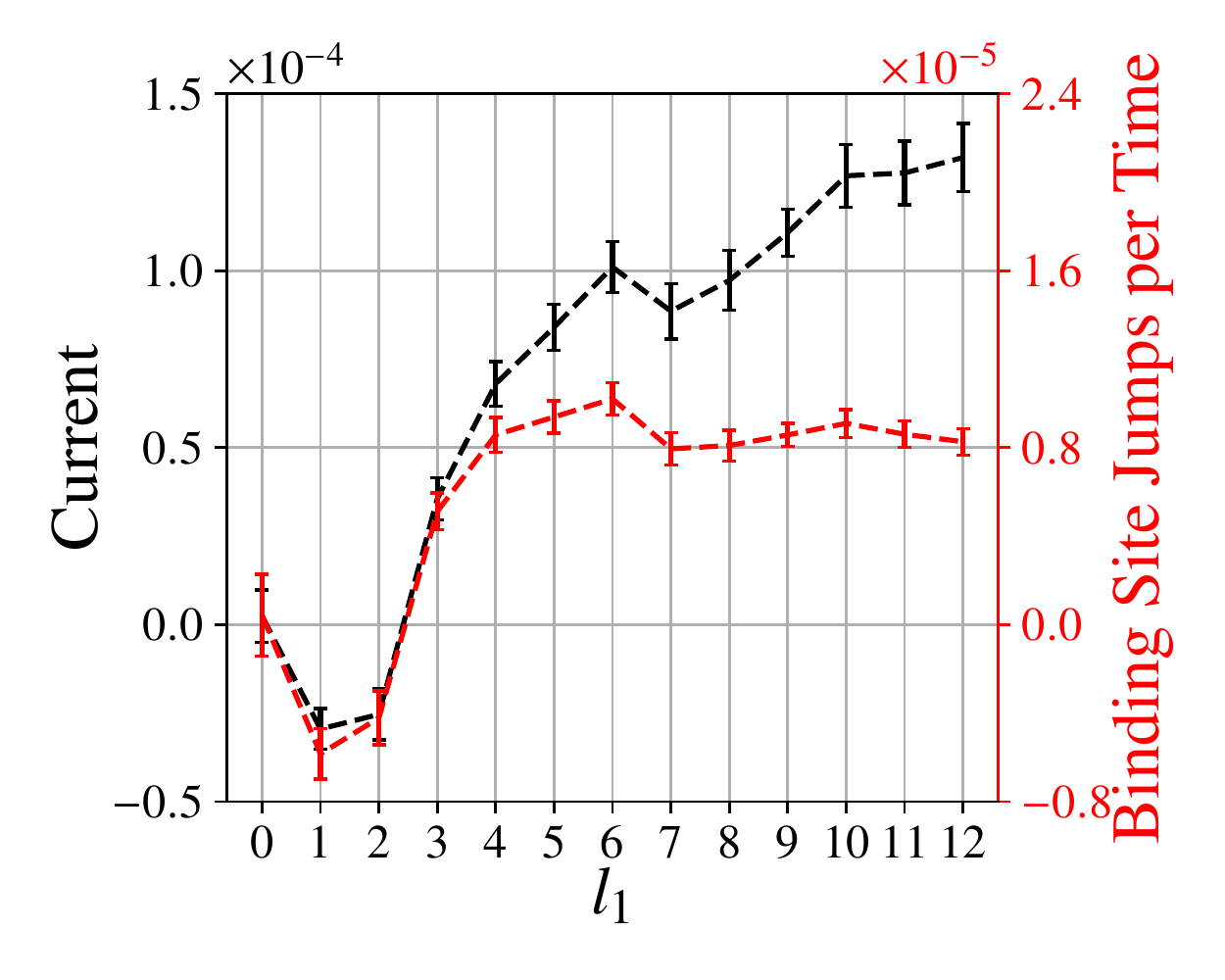}
\caption{
Current (left axis, black) and net number of jumps between binding sites made by the shuttling ring (right axis, red) as a function of \(l_{1}\) spacing in the linear motor for \(l_{2}=0\).
Data points and error bars are given by the mean and standard error across 100 independent simulations with \(2\times 10^8\) time steps of size \(\Delta t = 5 \times 10^{-3}\).
}
\label{fig:curr_vs_jumps}
\end{figure}

Fig.~\ref{fig:l1_l2} curiously shows that the linear motor's current (defined as the net number of track particles traversed by the shuttling ring per time) continues to increase as the \(l_{1}\) spacing increases.
This data is reproduced in black in Fig.~\ref{fig:curr_vs_jumps}.
If we instead examine the net number of jumps the shuttling ring makes between binding sites per time, shown in red in Fig.~\ref{fig:curr_vs_jumps}, we see that this value plateaus at large \(l_{1}\).
The increase in current at large \(l_{1}\) is then due to the large distance the shuttling ring must traverse when making a jump between binding sites, which occurs at a rate independent of \(l_{1}\).
This current increase is then not due to additional or changing steric effects at large \(l_{1}\), but simply the larger distance the ring must traverse between binding sites.

\section{Simulation Details}
\label{sec:details}
We used both circular (Fig.~\ref{fig:circular}b) and linear (Fig.~\ref{fig:l1_l2}a) tracks for the motor in this study.
The details and derivations of the methods and models used here are presented elsewhere and, unless otherwise noted, the model parameters correspond to ``Motor II'' from that study~\cite{albaugh2022simulating}.

The classical decomposition of a full tetrahedral cluster (FTC) into an empty tetrahedral cluster (ETC) and free central particle (C), depicted in Fig.~\ref{fig:circular}a, provides the driving force for motor operation.
The fuel was an FTC composed of a 4-particle tetrahedral cluster (blue) bound along its edges with harmonic potentials:
\begin{equation}
U_{\mathrm{harmonic}} (\mathbf{r}_{ij}) = \frac{1}{2} k_{ij} \mathbf{r}^{2}_{ij},
\end{equation}
where \( \mathbf{r}_{ij} \) is the distance vector between tetrahedron particles and \( k_{ij} \) is the harmonic force parameter.
A free C particle (red) was inserted into this tetrahedron.
To accommodate the C particle the cluster needs to deform from its equilibrium geometry, creating an entropically and energetically unfavorable state.
To escape from the tetrahedral cage, the C particle must deform the geometry even further, resulting in an even higher free energy transition state, causing FTC to be metastable.
Random thermal fluctuations can overcome this barrier, resulting in a transition to ETC + C.

The rotary motor with the circular track, depicted in Fig.~\ref{fig:circular}b, consisted of two interlocked rings.
Adjacent particles in each ring were connected with FENE bonds:
\begin{equation}
U_{\mathrm{FENE}} = -\frac{1}{2} k_{\mathrm{F},ij} r^{2}_{\mathrm{max},ij} \log{\left[ 1 - \left( \frac{| \mathbf{r}_{ij} |}{r_{\mathrm{max},ij}}\right)^{2} \right]},
\end{equation}
where \(\mathbf{r}_{ij}\) is the distance vector between neighboring ring particles \(i\) and \(j\) and \(k_{\mathrm{F},ij}\) and \(r_{\mathrm{max},ij}\) are force and maximum extension parameters, respectively.
Adjacent groups of three particles in the rings were also subject to an angular potential to maintain their circular geometry:
\begin{equation}
U_{\mathrm{angle}}(\theta_{ijk}) = \frac{1}{2}k_{\mathrm{A},ijk} \left(\theta_{ijk} - \theta_{0,ijk} \right)^{2},
\end{equation}
where \(\theta_{ijk}\) is the angle made by adjacent particles \(i\), \(j\), and \(k\), and \(k_{\mathrm{A},ijk}\) and \(\theta_{0,ijk}\) are force and equilibrium angle parameters, respectively.
The interlocked rings were free to diffuse around each other and within the simulation box.
The smaller shuttling ring (green) consisted of 12 particles and the larger ring consisted of 32 particles.
The linear motor, depicted in Fig.~\ref{fig:l1_l2}a, had the same 12-particle shuttling ring, but was then interlocked with a linear track of variable length that was fixed in place.
The larger ring and the linear track were composed of shuttling ring binding sites (orange), catalytic sites (white), and inert (black) particles.
Binding sites preferentially attracted the shuttling ring.
Inert particles had no attractive interactions, they were purely volume excluding.
The catalytic sites had attractive interactions with FTC particles that catalyze the FTC \(\to\) ETC + C reaction.

In general, attractive and repulsive interactions between any two particles in the system were given by a modified Lennard-Jones potential:
\begin{equation}
U_{\mathrm{LJ}}(\mathbf{r}_{ij}) = 4 \epsilon_{\mathrm{R},ij} \left( \frac{\sigma_{ij}}{|\mathbf{r}_{ij}|} \right)^{12} - 4 \epsilon_{\mathrm{A},ij} \left( \frac{\sigma_{ij}}{|\mathbf{r}_{ij}|} \right)^{6},
\label{eq:LJ}
\end{equation}
where \(\mathbf{r}_{ij}\) is the distance vector between particles \(i\) and \(j\), and \(\epsilon_{\mathrm{R},ij}\), \(\epsilon_{\mathrm{A},ij}\), and \(\sigma_{ij}\) are attractive energy, repulsive energy, and interaction size parameters, respectively.
This form of potential allowed us to change attraction and repulsion independently and every pairwise particle interaction was at least volume excluding (\( \epsilon_{\mathrm{R}} > 0\)), but not necessarily attractive (\( \epsilon_{\mathrm{A}} \ge 0\)).
Practically, we evaluated these pairwise interactions using a cell list~\cite{frenkel2001understanding}, where cell dimensions were no smaller than 4.25 on a side.
The LJ interactions were then switched smoothly to 0 over the range \(4.0 < |\mathbf{r}_{ij}| \le 4.25\) to ensure that interactions did not extend beyond nearest neighbor cells.
The switched LJ potential was
\begin{equation}
U_{\mathrm{LJ}}^{\mathrm{switch}} =  
\begin{cases}
U_{\mathrm{LJ}}(\mathbf{r}_{ij}) & \text{if }  |\mathbf{r}_{ij}| \le 4.0\\
U_{\mathrm{LJ}}(\mathbf{r}_{ij}) (1 + \lambda^{2}(2 \lambda - 3) ) & \text{if }  4.0 < |\mathbf{r}_{ij}| \le 4.25\\
0 & \text{if } |\mathbf{r}_{ij}| > 4.25,
\end{cases}
\label{eq:switch}
\end{equation}
where \(\lambda = (|\mathbf{r}_{ij}| - r_{\mathrm{cut}} + r_{\mathrm{switch}})/ r_{\mathrm{switch}}\), \(r_{\mathrm{cut}}=4.25\), and \(r_{\mathrm{switch}}=0.25\).

Most force field parameters for interactions between different particle types (\(k_{\mathrm{F},ij}\), \(r_{\mathrm{max},ij}\), \(k_{\mathrm{A},ijk}\), \(\theta_{0,ijk}\), \( k_{ij} \), \(\epsilon_{\mathrm{R},ij}\), \(\epsilon_{\mathrm{A},ij}\), and \(\sigma_{ij}\)) were identical to those of the ``Motor II'' parameterization in Ref.~\cite{albaugh2022simulating}.
In the present study we introduced new attraction and repulsion LJ parameters between the catalytic site and the FTC particles.
The previous study used three different types of catalytic particles (CAT1, CAT2, and CAT3) arranged CAT2-CAT1-CAT3 at each 3-particle catalytic site.
Additionally, there were four distinct types of tetrahedron particles (TET1, TET2, TET3, and TET4).  
The specificity of these reactions was what led to the effective catalysis, but in the previous study this meant that the fuel-catalyst interactions were not symmetric.
In the present study we sacrificed some specificity to symmetrize the catalyst, in order to make sure that catalyst asymmetry was not a confounding factor in the current reversal.
The catalyst here consisted of only two types of catalyst particles (CAT2 and CAT1) arranged in a CAT2-CAT1-CAT2 configuration for each 3-particle catalytic site.
The symmetrized FTC-catalyst interactions we used are given in Table~\ref{tab:tab1}.

\begin{table}
\centering
\caption{Catalytic site-FTC modified Lennard-Jones interaction parameters.}
\begin{tabular}{c|c|c}
Particle Pair & \(\epsilon_{\mathrm{A}}\) & \(\epsilon_{\mathrm{R}}\) \\
\hline
\midrule
CAT1-TET1 & 0.732877 & 0.888468  \\
CAT1-TET2 & 0.573471 & 1.215096  \\
CAT1-TET3 & 0.221362 & 1.424995 \\
CAT1-TET4 & 0.511311 & 0.216518 \\
CAT1-CENT & 3.922903 & 2.798031 \\
CAT2-TET1 & 1.457497 & 1.147790  \\
CAT2-TET2 & 1.253230 & 1.440685  \\
CAT2-TET3 & 0.928847 & 2.063941 \\
CAT2-TET4 & 1.567248 & 3.788327 \\
CAT2-CENT & 1.196286 & 1.421851 \\
\bottomrule
\end{tabular}
\label{tab:tab1}
\end{table}

The dynamics of the particles (excluding those of the linear track, which was fixed in place) were driven by a Langevin equation of motion.
For a particle \(i\) this equation is
\begin{equation}
\begin{aligned}
\dot{\mathbf{r}}_{i} & = \frac{\mathbf{p}_{i}}{m_{i}}
\\
\dot{\mathbf{p}}_{i} &= - \frac{\partial U(\mathbf{r})}{\partial \mathbf{r}_{i}} - \frac{\gamma}{m_{i}} \mathbf{p}_{i} + \boldsymbol{\xi}_{i},
\end{aligned}
\end{equation}
where the potential energy \(U\) is a function of all positions \(\mathbf{r}\), \(m_{i}\) is the particle's mass, \(\gamma\) is the friction coefficient and \(\boldsymbol{\xi}\) is a white noise satisfying \(\left<\boldsymbol{\xi}_{i}\right> = \mathbf{0}\) and \(\left<\boldsymbol{\xi}_{i}(t) \boldsymbol{\xi}_{j}(t')\right> = 2 \gamma k_{\rm B} T \delta(t - t') \delta_{ij} \mathbf{I}\) at temperature \(T\), where \(\mathbf{I}\) is the identity matrix.
This equation was numerically integrated with the VRORV scheme~\cite{fass2018quantifying}:
\begin{equation}
\begin{aligned}
\mathbf{p}_{i}& \left(t+\frac{1}{4}\Delta t \right) = \mathbf{p}_{i} \left(t \right) - \frac{\Delta t}{2} \frac{ \partial U \left( \mathbf{r} \left( t \right) \right)} {\partial \mathbf{r}_{i} \left(t \right) }
\\
\mathbf{r}_{i}& \left( t+\frac{1}{2}\Delta t \right) = \mathbf{r}_{i}\left(t \right) + \frac{\Delta t}{2 m_{i}} \mathbf{p}_{i} \left( t+\frac{1}{4}\Delta t \right)
\\
\mathbf{p}_{i} &\left( t+\frac{3}{4}\Delta t \right) = \\
&e^{\frac{-m_{i} \Delta t}{\gamma}}\mathbf{p}_{i} \left( t+\frac{1}{4}\Delta t \right) + \sqrt{m_{i} \frac{1- e^{\frac{-2 m_{i} \Delta t}{\gamma}} }{\beta }} \boldsymbol{\eta}_{i} \left( t+\frac{3}{4}\Delta t \right)
\\
\mathbf{r}_{i} &\left( t+\Delta t \right) = \mathbf{r}_{i} \left( t+\frac{1}{2}\Delta t \right) + \frac{\Delta t}{2 m_{i}}\mathbf{p}_{i}\left( t+\frac{3}{4}\Delta t \right)
\\
\mathbf{p}_{i}& \left(t+\Delta t \right) = \mathbf{p}_{i} \left(t+\frac{3}{4}\Delta t \right) - \frac{\Delta t}{2} \frac{ \partial U \left( \mathbf{r} \left(t+\Delta t \right) \right)}{\partial \mathbf{r}_{i} \left(t + \Delta t \right) },
\end{aligned}
\end{equation}
where \(\boldsymbol{\eta}_{i}\) is a random vector with components drawn from a zero mean, unity variance normal distribution and \(\beta = (k_{\mathrm{B}} T)^{-1}\).
The VRORV acronym reflects the order that dynamic variables are updated in each time step.
First the velocity or momentum (V) is updated, then the position (R), then an Ornstein–Uhlenbeck (O) random process updates the velocity, then another position (R) update, and a final velocity or momentum (V) update.
This procedure was repeated for \(N_{\mathrm{steps}}\) total time steps per simulation.
Unless otherwise noted, the temperature was \(T=0.5\), the time step was \(\Delta t = 0.005\), and the friction was set to \(\gamma=0.5\).
The simulations and models were non-dimensionalized with the energy scale set by the repulsion between inert (black) particles \(\epsilon_{\mathrm{R,} \mathrm{INERT-INERT}}\), the length scale set by the LJ radius of the inert particles \(\sigma_{\mathrm{INERT},\mathrm{INERT}}\), and the mass scale set by the mass of the inert particles \(m_{\mathrm{INERT}}\).
The Boltzmann constant \(k_{\mathrm{B}}\) was set to 1.
All data are presented in terms of these reduced units.

For the rotary motor, the simulation cell consisted of two concentric cubic boxes with edge lengths \(L_{\mathrm{inner}}=30\) and \(L_{\mathrm{outer}}=34\).
The motor rings were confined to the inner simulation box with a Lennard-Jones wall potential:
\begin{equation}
\begin{aligned}
U_{\mathrm{wall}}(\mathbf{r}_{i}) = 4 \epsilon_{\mathrm{wall}} \sum_{\alpha=x,y,z} \Bigg[ &\left( \frac{\sigma_{\mathrm{wall}}}{r_{\alpha,i} - \frac{1}{2} L_{\mathrm{inner}}} \right)^{12}\\
&+\left( \frac{\sigma_{\mathrm{wall}}}{r_{\alpha,i} + \frac{1}{2} L_{\mathrm{inner}}} \right)^{12} \Bigg],
\end{aligned}
\end{equation}
where \(\mathbf{r}_{i}\) is a motor particle, \(\epsilon_{\mathrm{wall}}=1\) and \(\sigma_{\mathrm{wall}}=1\) are wall energy and size parameters, respectively.
Particles of the FTC, ETC, and C species did not interact with the wall potential, allowing them to diffuse freely between the inner and outer boxes.
The faces of the outer box were periodic, allowing the species to exit one face of the box and immediately re-enter through the opposite face.
The box dimensions equate to an inner volume (accessible to the motor) of \(V_{\mathrm{inner}}=27000.0\), a total volume of \(V_{\mathrm{outer}}=39304.0\), and a volume available for GCMC moves of \(V_{\mathrm{GCMC}}=V_{\mathrm{outer}}-V_{\mathrm{inner}}=12304.0\).

For the linear motor, the track was fixed in position, centered in the \(yz\)-plane at (\(y=0\), \(z=0\)) with a spacing of \(d=0.992\) between track particles, the approximate equilibrium distance between particles in the large ring of the rotary motor.
As depicted in Fig.~\ref{fig:l1_l2}, the inner and outer simulation boxes were no longer necessarily cubic and they shared boundaries in the \(yz\)-plane.
The shared inner and outer box lengths in the \(x\)-direction \(L_{x}\) were determined by the size of linear motor itself, \(L_{x} = d N_{\mathrm{motif}}(l_{1} + l_{2} + 4)\) (where there were \(l_{1} + l_{2}\) inert particles, 3 catalytic particles, and 1 binding particle per repeat unit).
The shared end faces were then both periodic allowing FTC, ETC, C, and the shuttling ring to cross.
In this way the shuttling ring diffused on an effectively infinite linear track.
The remaining \(y\) and \(z\) dimensions of the inner and outer box were set so that the \(y\) and \(z\) dimensions were equivalent and that \(V_{\mathrm{inner}}=27000.0\) and \(V_{\mathrm{GCMC}}=12304.0\), for consistency with the rotary motor.
By keeping these volumes consistent across all simulations and all configurations, we ensured that concentrations, free energies, and shifted chemical potentials were equivalent across all simulations.
We calculated \(L_{y,\mathrm{inner}} = L_{z,\mathrm{inner}} = \sqrt{V_{\mathrm{inner}}/L_{x}}\) and \(L_{y,\mathrm{outer}} = L_{z,\mathrm{outer}} = \sqrt{(V_{\mathrm{inner}} + V_{\mathrm{GCMC}}) /L_{x}}\).
For completeness we still implemented a wall potential in the \(y\) and \(z\) directions applied to the shuttling ring (although the shuttling ring was mechanically interlocked with the central track, making it impossible to approach the wall):
\begin{equation}
\begin{aligned}
U_{\mathrm{wall}}(\mathbf{r}_{i}) = 4 \epsilon_{\mathrm{wall}} \sum_{\alpha=y,z} \Bigg[ &\left( \frac{\sigma_{\mathrm{wall}}}{r_{\alpha,i} - \frac{1}{2} L_{\alpha, \mathrm{inner}}} \right)^{12}\\
&+\left( \frac{\sigma_{\mathrm{wall}}}{r_{\alpha,i} + \frac{1}{2} L_{\alpha, \mathrm{inner}}} \right)^{12} \Bigg],
\end{aligned}
\end{equation}
where \(\mathbf{r}_{i}\) is a shuttling ring particle, and \(\epsilon_{\mathrm{wall}}=1\) and \(\sigma_{\mathrm{wall}}=1\), as before.

Periodically we performed GCMC insertion and deletion moves of the FTC, ETC, and C species in the space between the inner and outer boxes~\cite{frenkel2001understanding}.
GCMC moves were attempted every 100 Langevin time steps.
By keeping the motor spatially separated from the GCMC moves, we ensured that these instantaneous changes did not directly affect the motor's dynamics.
For each GCMC move we selected randomly between the six possible moves (insertion or removal of FTC, ETC, and C) with uniform probability.
For insertion moves, the momenta of the inserted species were drawn from a Boltzmann distribution and for FTC and ETC their positions were drawn from a pre-sampled canonical library of \(10^{4}\) configurations~\cite{gupta2000grand,chempath2003two}.
The subsequent acceptance probabilities for GCMC insertion moves and removal moves of species \(i\), respectively, are:
\begin{equation}
P^{\mathrm{acc}}_{i,\mathrm{insertion}}(\mathbf{r}^{\prime},\mathbf{p}^{\prime} \to \mathbf{r},\mathbf{p}) = \mathrm{min} \left[1, \frac{1}{N_{i}(\mathbf{r}^{\prime})} e^{-\beta (U(\mathbf{r}^{\prime}) - U(\mathbf{r}) - U_{i}^{0} - \mu_{i}^{\prime}) } \right]
\end{equation}
and
\begin{equation}
P^{\mathrm{acc}}_{i,\mathrm{removal}}(\mathbf{r}^{\prime},\mathbf{p}^{\prime} \to \mathbf{r},\mathbf{p}) = \mathrm{min} \left[1, N_{i}(\mathbf{r}) e^{-\beta (U(\mathbf{r}^{\prime}) - U(\mathbf{r}) + U_{i}^{0} + \mu_{i}^{\prime}) } \right],
\end{equation}
where \(\mathbf{r}\) and \(\mathbf{p}\) are the positions and momenta of the initial configuration, \(\mathbf{r}^{\prime}\) and \(\mathbf{p}^{\prime}\) are the positions and momenta of the trial configurations, \(N_{i}\) is the number of molecules of species \(i\), \(U(\mathbf{r})\) is the potential energy of the initial configuration, \(U(\mathbf{r}^{\prime})\) is the potential energy of the trial configuration, \(U_{i}^{0}\) is the internal potential energy of the isolated molecule being inserted or removed, and \(\mu_{i}^{\prime}\) is the relative chemical potential.
The relative chemical potential is the true chemical potential less the Helmholtz free energy of a single isolated species of \(i\) in the same volume, \(\mu_{i}^{\prime} = \mu_{i} - A_{i}^{0}\).
Unless otherwise noted, these external relative chemical potentials were set to \(\mu_{\mathrm{FTC}}^{\prime}=0.5\), \(\mu_{\mathrm{ETC}}^{\prime}=-10.0\), and \(\mu_{\mathrm{C}}^{\prime}=-10.0\). 
These values ensured that FTC tended to be inserted while ETC and C tended to be removed, resulting in a nonequilibrium concentration of FTC and practically no free ETC and C in the simulation cell.

\end{document}